\documentclass[12pt]{article}
\usepackage{arxiv}
\usepackage{amssymb}
\usepackage{nicefrac}
\usepackage[utf8]{inputenc} 
\usepackage[T1]{fontenc}    
\usepackage{lipsum}
\usepackage{parskip}
\usepackage{stmaryrd}
\usepackage{mathrsfs}
\usepackage[colorlinks]{hyperref}
\usepackage{graphicx}
\usepackage{amsmath}
\usepackage{amssymb}
\usepackage[dvipsnames]{xcolor}
\usepackage{subfig}
\usepackage{comment}
\usepackage{appendix}
\usepackage{here}
\usepackage{gensymb}
\usepackage{mathrsfs}
\usepackage{ctable}
\usepackage{bm}
\usepackage{multirow}
\usepackage{textcomp}
\usepackage{siunitx}[=v2]
\DeclareSIUnit{\wtpercent}{wt\%}
\newcommand{\murm}{%
  \ifmmode
    \mathchoice
        {\hbox{\normalsize\textmu}}
        {\hbox{\normalsize\textmu}}
        {\hbox{\scriptsize\textmu}}
        {\hbox{\tiny\textmu}}%
  \else
    \textmu
  \fi
}

\title{A level-set formulation to simulate diffusive solid/solid phase transformation in polycrystalline metallic materials - Application to austenite decomposition in steels}

\author{{N.~Chandrappa}\\
Mines Paris, PSL University\\
Centre for material forming (CEMEF), UMR CNRS\\
06904 Sophia Antipolis, France\\
\texttt{nitish.chandrappa@mines-paristech.fr}\\
\And
{M.~Bernacki}\\
Mines Paris, PSL University\\
Centre for material forming (CEMEF), UMR CNRS\\
06904 Sophia Antipolis, France\\
\texttt{marc.bernacki@mines-paristech.fr}
}

\renewcommand{\shorttitle}{}

\begin{document}
\maketitle
\begin{abstract}
Numerous full-field numerical methods exist concerning the digital description of polycrystalline materials and the modeling of their evolution during thermomechanical treatments. However, these strategies are globally dedicated to the modeling of recrystallization and grain growth for single-phase materials, or to the modeling of phase transformations without considering recrystallization and related phenomena. A generalized numerical framework capable of making predictions in a multi-phase polycrystalline context while respecting the concomitance of the different microstructural mechanisms is thus of prime interest. 
A novel finite element level-set based full-field numerical formulation is proposed to principally simulate diffusive solid-solid phase transformation at the mesoscopic scale in the context of two-phase metallic alloys. A global kinetic framework, capable of accounting for other concomitant mechanisms such as recrystallization and grain growth is considered in this numerical model. The proposed numerical framework is shown to be promising through a couple of illustrative 1D and 2D test cases in the context of austenite decomposition in steels and compared with ThermoCalc estimations.     
\end{abstract}



\section{Introduction}
\label{intro}
It has been well established that the metallic material properties have a direct correlation with the underlying microstructure. When a material is subjected to thermomechanical treatments (TMTs) in the context of metal forming, several microstructural changes \cite{martin1997stability} could occur in the form of recovery, recrystallization (ReX), grain growth (GG), phase transformation (PT) which in-turn modify the material's macroscopic properties. Phase transformation at the solid-state involves crystallographic changes in the parent phase through rearrangement of the lattice structure to form a different, more stable product phase at the same solid state. PT can be either displacive or diffusive. Displacive transformation \cite{jamesDispPT} is characterized by the spontaneous, coherent, and cooperative movement of atoms across distances that are typically smaller than one nearest neighbor spacing. Diffusive transformation \cite{GAMSJAGER200292} involves gradual reorganization of the lattice through short and long-range diffusion of atoms. Two basic mechanisms drive diffusive PT: (i) the diffusion of solutes across the phase interfaces and in the bulk of the grains, resulting in a change in chemical composition, and (ii) the interface migration resulting in the lattice rearrangement or structural changes. PT plays a critical role in producing diverse materials with varying microstructural features during TMTs. Considering the large-scale use of metallurgical products in various strategic industries (nuclear, aerospace, automotive, oil \& gas, defense, and renewable energies et cetera), a comprehensive understanding and modeling of microstructural mechanisms during TMTs are of prime importance. There is then a growing demand to develop more physically realistic numerical models capable of precisely predicting the microstructural evolution and in turn determine the in-service material performances.

To model microstructural evolution, depending on the level of description desired, we broadly have three main modeling approaches: mean-field modeling and full-field modeling at the mesoscopic scale, and molecular dynamics. Mean-field models (MFM) \cite{Montheillet2009, Maire2018, perez2008, seret2020} are based on an averaged description of the microstructure by considering grains or precipitates as spherical entities, and involving statistical evolution related to different characteristics (grain size, precipitate size, dislocation density, ...). MFM are then computationally efficient but do not involve precise modeling of the topological changes. Advances in computational resources have paved the way for more intricate models (such as atomistic and full-field mesoscopic models) capable of explicitly reproducing the microstructural evolution. Molecular dynamics \cite{jin2006atomic, mishin2010atomistic} approaches consider the basic building blocks of material, atoms, as the smallest entity. Such models provide a profound description of the involved mechanisms but also require large computational resources. Thus, these models are often considered to analyze or quantify certain characteristics over a localized region of the microstructure limited to a few interfaces. By simplifying the interface description and approximating the interface properties and kinetics, the so-called mesoscopic full-field models (FFM) \cite{janssens2010computational}, are based on an explicit description of the microstructure topology at the polycrystalline scale by typically considering few thousand to few ten thousand grains in 2D or 3D. FFM have demonstrated an exciting potential to simulate a wide range of microstructural evolution such as the precise modeling of ReX in dynamic (DRX) or post-dynamic (PDRX) conditions, GG, diffusive solid-solid phase transformation (DSSPT), spheroidization and sintering. In the context of microstructural evolution, FFM mainly comprise the following numerical methods: Monte Carlo (MC) Potts \cite{ROLLETT1989627}, Cellular Automata (CA) \cite{raabe2002}, Phase-Field (PF) or Multi Phase-Field (MPF) \cite{Steinbach2009, KRILLIII20023059, steinbach2020}, Front-Tracking \cite{florez:hal-03030705, florez:hal-03425706}/ Vertex methods \cite{Barrales2008}, and Level-Set (LS) models \cite{Merriman1994MotionOM, bernacki:hal-00509731}.    

In the context of DSSPT, phase-field methods (PFM) are popular and extensively used. The thermodynamic consistency and the ability to model arbitrary complex morphological changes without any presumption on their shape or mutual distribution make PFM a powerful and an attractive tool. The early works of Wheeler et al. \cite{wheeler1992}, Steinbach et al. \cite{STEINBACH1996135, TIADEN199873} on solidification using PFM provided some of the mathematical foundations of phase-field modeling for multi-component, multi-phase systems involving solute diffusion. Yeon et al. \cite{Yeon2001} presented one of the first phase-field simulations of DSSPT, where austenite-ferrite transitions in the Fe-Mn-C system were modeled under para-equilibrium \cite{hillert2004definitions} assumptions. Pariser et al. \cite{Pariser2001} studied the phase transformation behavior in ULC (Ultra Low Carbon) and IF (Interstitial free) grade steels using the well-known MICRESS software \cite{micress} based on the multi-component, multi phase-field method. Huang et al. \cite{Huang2006} performed 2D PF simulations for $\gamma \rightarrow \alpha$ transformation in low carbon steels by considering an arbitrary number of grains at a large spatial scale. The P.h.D. works of Mecozzi along with Militzer et al. \cite{Mecozzi2007PhaseFM, Militzer2006} were dedicated to the first 3D simulations of DSSPT for $\gamma \rightarrow \alpha$ transformation in a Fe-Mn-C system. In addition to austenite decomposition in steels, there have been works dedicated to other alloyed materials also. 1D PF simulations for phase transformation in aluminum alloys have been studied in \cite{kovacevic2006}. Malik et al. \cite{Malik2017} have used 2D PFM to simulate the formation and growth of $\sigma-$phase precipitates in a super duplex stainless steel alloy. Some other works are based on MC \cite{TONG20051485} and CA methods \cite{varma2001, Lan2004}. In most of these reported works, GG aspects were either completely neglected or only the GG of the product phase was accounted for while ignoring that of the parent phase. 

Moreover, in the context of industrial processes where high plastic deformation can be achieved, none of the existing approaches provide easily an appropriate framework to perform simulations of DRX concomitant with phase transformation in multi-phase materials. On the other hand, level-sets (LS) have been successfully used to simulate DRX \cite{MairePhD} and GG phenomena \cite{bernacki:hal-00577039, FAUSTY2018578, alvarado2021} for single-phase materials. So, in the current state of the art, most of the numerical predictions are dedicated to single-phase microstructural evolution, or only based on phase transformation without taking into account other phenomena such as ReX or GG. Such numerical approaches can then be insufficient when complex thermomechanical treatments with large temperature ranges are investigated. Thus, there is a need for a generalized numerical framework capable of making predictions of DSSPT, DRX, and GG in a multi-phase polycrystalline context \cite{liss2009situ}. So, the perspective of this work is to explore the potential of the LS method for the modeling of DSSPT. We thus propose a global finite-element (FE) LS formalism capable of simulating diffusive phase transformation and ReX in the context of large plastic deformation for multi-phase polycrystalline materials by considering the driving pressures acting on grain and phase interfaces.  

The proposed LS based numerical formulation in the context of austenite decomposition (austenite to ferrite phase transformation) is described in section \ref{numform}. In section \ref{resdisc}, a couple of representative illustrations are used to demonstrate the potential of the numerical model. Finally, in section \ref{concl}, we discuss the key remarks of the proposed approach and also some perspectives for future work.

\section{Numerical formulation: Diffusive phase transformation modeling}
\label{numform}

DSSPT modeling at the mesoscopic scale typically involves two governing equations: a diffusion equation that governs the partitioning of solute atoms (such as carbon) across different phases, and another governing equation that takes care of the resulting interface network migration. As mentioned earlier, our interest is to use a global LS formalism to simulate the considered phenomena. Classically, the diffusion equation could be resolved within a LS framework. However, due to the presence of material discontinuities across the phase interfaces, the sharp interface approach considered in the LS framework enforces the explicit consideration of interface jump conditions during the resolution of the diffusion equation. This demands explicit localization of the interface at each instant to treat numerically the necessary jump conditions. Thus, to avoid this cumbersome step, we propose to consider a diffuse interface hypothesis across the phase interfaces during the resolution of the diffusion equation. In other words, we represent and migrate the multi-phase grain interface network using a LS description while resolving a global diffusion equation based on a diffusive interface assumption for the phase interfaces. The diffuse interface description is realized using a phase-field like function which ensures that any material discontinuities across the interface are naturally smoothened. This enables us to resolve a single diffusion equation in the whole computational domain without the need for any interface jump conditions.  

\begin{figure}[!htb]
	\centering
	\captionsetup{justification=centering}
	\includegraphics[width=12cm]{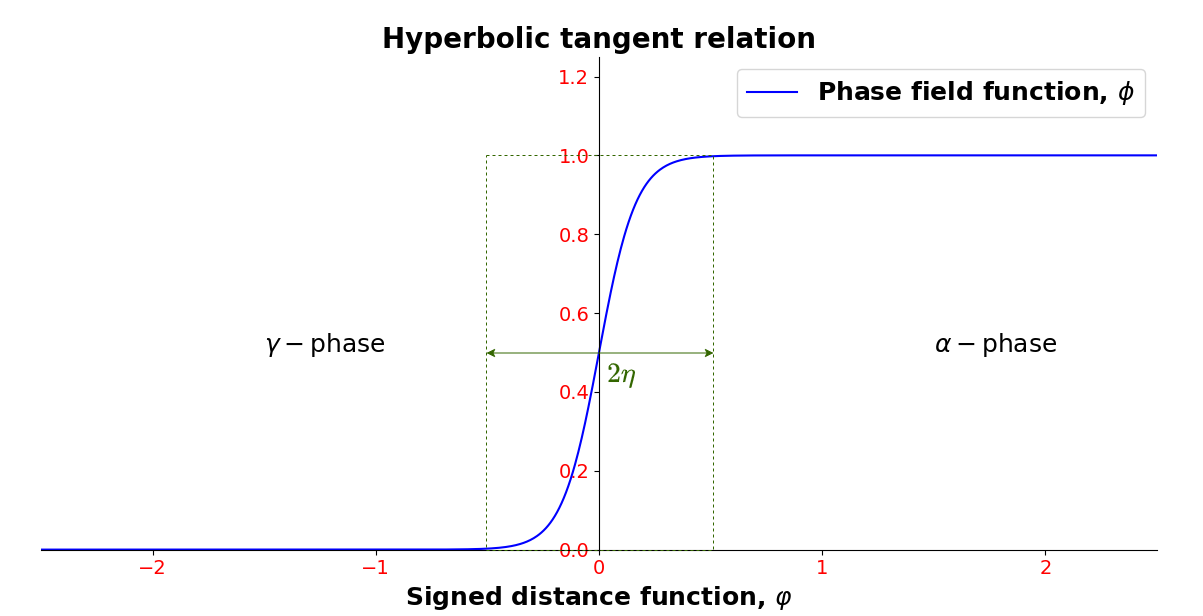}
	\caption{Hyperbolic tangent relation yielding the diffuse phase interface between phases $\alpha$ and $\gamma$}
	\label{hyptanplot}
\end{figure}

This transition to a diffuse interface description is established, thanks to a hyperbolic tangent relation \cite{Mecozzi2007PhaseFM, OLSSON2005225} between a phase-field like function ($\phi$) and a signed distance LS function ($\varphi$) of the following form: 
\begin{equation}
\phi = \frac{1}{2}tanh\left(\frac{3\varphi}{\eta}\right)+\frac{1}{2},
\label{tanhf}
\end{equation}where $\eta$ is a diffuse interface thickness parameter. In the following, we shall refer this function ($\phi$) yielding the diffuse interface as the phase-field function. Fig. \ref{hyptanplot} illustrates the trend of this function in a 1D context.

\subsection{Solute partitioning:}
Let us assume the solute concentration in the parent phase (austenite, $\gamma$) and the product phase (ferrite, $\alpha$) to be $C_\gamma$ and $C_\alpha$ respectively. After having established a diffuse interface description, the total carbon concentration field ($C$) can be expressed as a continuous variable:
\begin{equation}
C = \phi C_\alpha + (1-\phi)C_\gamma
\label{cmix}.
\end{equation}
Likewise, we then assume continuity of the solute fluxes of each phase ($\bm{J}_\gamma, \bm{J}_\alpha$) weighted by the phase-field variable across the phase interface:
\begin{equation}
\bm{J} = \phi \bm{J}_{\alpha} + (1-\phi)\bm{J}_{\gamma}.
\label{fluxmix}
\end{equation}
The diffuse phase interface is assumed to be composed of a mixture of the two phases. A constant concentration ratio is imposed between the phases, such that the redistribution of the solute atoms between them at the interface respects a partitioning ratio ($k$) equal to that at the equilibrium:
\begin{equation}
k = \frac{C_\alpha}{C_\gamma} =\frac{C_\alpha^{eq}}{C_\gamma^{eq}}
\label{keq},
\end{equation}where $C_\alpha^{eq}$ and $C_\gamma^{eq}$ are the equilibrium concentrations of $\alpha$ and $\gamma$ phases respectively at temperature $T$.

Following Fick's laws of diffusion, the diffusion equation for carbon partitioning can be expressed as:
\begin{equation*}
\frac{\partial C}{\partial t} = -\bm{\nabla}\cdot\bm{J} = -\bm{\nabla}\cdot\left[\phi \bm{J}_{\alpha} + (1-\phi)\bm{J}_{\gamma}\right],
\end{equation*}
with, 
\begin{equation*}
\bm{J}_{\alpha} = -D_\alpha^C \bm{\nabla} C_\alpha; \quad  \bm{J}_{\gamma}=-D_\gamma^C \bm{\nabla} C_\gamma.
\end{equation*}
We then obtain,
\begin{equation}
\frac{\partial C}{\partial t} = \bm{\nabla}\cdot\left[\phi D^C_\alpha \bm{\nabla} C_\alpha + (1-\phi)D^C_\gamma \bm{\nabla} C_\gamma \right],
\label{diffmixeqn}
\end{equation}where $D^C_\alpha$ and $D^C_\gamma$ represent the diffusivity of the carbon element in ferrite and austenite phases respectively.

Invoking eqs.\eqref{cmix} and \eqref{keq} in eq.\eqref{diffmixeqn}, a modified carbon diffusion equation \cite{Mecozzi2007PhaseFM, TIADEN199873} is obtained:
\begin{equation}
\frac{\partial C}{\partial t} = \bm{\nabla}\cdot\left\{D^*(\phi)\left[\bm{\nabla} C -\frac{C(k-1)}{1+\phi(k-1)}\bm{\nabla}\phi \right]\right\},
\label{diffmixform}
\end{equation}where $D^*(\phi)$ is called "mixed diffusivity" and is defined as,
\begin{equation*}
D^*(\phi) = \frac{D_\gamma^C + \phi(kD^C_\alpha - D^C_\gamma)}{1+\phi(k-1)}.
\label{mixeddiffus}
\end{equation*}

With further simplifications, the above eq.\eqref{diffmixform} can be transformed into a Convective-Diffusive-Reactive (CDR) form as follows:

\begin{equation*}
\frac{\partial C}{\partial t} = \bm{\nabla}\cdot\left[D^*(\phi)\bm{\nabla}C - C\bm{A}(\phi)\right] 
\end{equation*}

\begin{equation}
\frac{\partial C}{\partial t} +  \left(\bm{A}-\bm{\nabla}D^*\right)\cdot\bm{\nabla}C - D^*\Delta C + RC = 0,
\label{CDReqn} 
\end{equation} where,
\begin{equation*}
\bm{A}(\phi)= \frac{D^*(\phi)(k-1)}{1+\phi(k-1)}\bm{\nabla}\phi, \quad and \quad R=\bm{\nabla}\cdot\bm{A}.
\end{equation*}


Let $\psi \in H^1(\Omega)$ be a test function, the FE weak formulation of eq.\eqref{CDReqn} can be written as follows:

\begin{equation}
\int_\Omega \frac{\partial C}{\partial t}\psi d\Omega  +\int_\Omega\left(\bm{A}-\bm{\nabla}D^*\right)\cdot\bm{\nabla}C\psi d\Omega - \int_\Omega D^*\Delta C\psi d\Omega + \int_\Omega RC\psi d\Omega = 0.
\label{wfstep1}
\end{equation}

Since we assume no influx or outflux of solute atoms into or from the domain respectively, solute mass is conserved at any instant. Thus imposing pure Neumann boundary conditions on the boundaries of the computational domain ($\bm{\nabla}C\cdot\bm{n}\vert_{\partial \Omega}=0$), and applying the divergence theorem, we have:
\begin{equation*}
\int_\Omega D^*\Delta C\psi d\Omega = \int_{\partial \Omega} \psi D^*\bm{\nabla}C\cdot\bm{n} dS - \int_\Omega \bm{\nabla}(D^*\psi)\cdot \bm{\nabla}C d\Omega = - \int_\Omega \bm{\nabla}(D^*\psi)\cdot \bm{\nabla}C d\Omega.
\end{equation*}

Substituting the above term in eq.\eqref{wfstep1}, and after simplification, we get:
\begin{equation}
\int_\Omega \frac{\partial C}{\partial t}\psi d\Omega  +\int_\Omega\bm{A}\cdot\bm{\nabla}C\psi d\Omega + \int_\Omega D^*\bm{\nabla}\psi\cdot \bm{\nabla}C d\Omega + \int_\Omega RC\psi d\Omega = 0.
\label{diffWF} 
\end{equation}
It can be highlighted that compared to the strong formulation in eq.\eqref{CDReqn}, the gradient of the mixed diffusivity term ($\bm{\nabla} D^*$) vanishes in the weak formulation. In terms of numerical stability, this is of great interest considering the abrupt evolution of this term across a phase interface.

\subsection{Interface migration}
To govern the motion of the multi-phase grain interface network, we revert to the LS description for the interfaces. Considering the interface of interest, $\Gamma$, of a closed domain $G$, our LS is classically initialized as a signed Euclidean distance function to $\Gamma$ such that the zero isovalue of this function localizes the interface $\Gamma$:  
\begin{equation*}
\begin{cases}
\varphi(\bm{x})=\pm d(\bm{x}, \Gamma), \quad \bm{x}\in \Omega \\
\Gamma = \partial G= \left\{\bm{x} \in \Omega, \varphi(\bm{x})=0\right\}
\end{cases}.
\label{signLS}
\end{equation*}
In the following, $\varphi$ will be assumed positive inside $G$ and negative outside. Considering $\bm{v}$, the kinetic of the $\Gamma$ interface, at any time, $\Gamma(t)$ can be obtained by solving the following convective equation \cite{sethian1999level}:
\begin{equation}
\begin{cases}
\frac{\partial \varphi_i}{\partial t} + \bm{v}\cdot\bm{\nabla}\varphi_i=0 \\
\varphi_i(\bm{x}, t=0) = \varphi_i^0(\bm{x})
\end{cases}	\forall i \in \{1,2,...,N_{LS}\},
\label{lstranseqn}
\end{equation}where $N_{LS}$ is the number of active level-set functions used to represent different grains of different phases in the microstructure. The classical approach is to consider one level-set function per grain. However, such an approach is totally inefficient when a large number of grains is considered. Thus, a grain coloration/ re-coloration scheme proposed in \cite{Scholtes2015} is used to limit the number of LS functions required. The re-coloration scheme ensures that there are no instances of numerical coalescence of two or more grains close to each other during the migration of the grain and phase boundary network.

In the context of microstructural evolution at the mesoscopic scale, the velocity field, $\bm{v}$ is assumed to be a product of the interface mobility ($\mu$) and the different driving pressures P, describing the involved phenomena \cite{rollett2017recrystallization, christian1975theory}:
\begin{equation}
\bm{v} = \mu P\bm{n},
\label{ssptkin}
\end{equation}where $\bm{n}$ is the outward unit normal vector to the considered interface.

Typically, in the context of hot metal forming, the main driving pressures leading to phase and grain evolution are: (i) $P_c=\Delta G$, which is the difference in Gibbs free energy between different phases and is the principal component responsible for phase transformation, (ii) $P_s=\llbracket E \rrbracket$, which is the jump in stored energy due to plastic deformation, responsible for recrystallization phenomenon, and (iii) $P_\kappa=-\kappa\sigma$, where $\kappa$ is the trace of the curvature tensor of the interface and $\sigma$ is the interfacial energy. This pressure corresponds to the capillarity effects through the minimization of surface energy due to the presence of grain and phase interfaces (well known as the Gibbs-Thomson effect), and is responsible for grain growth phenomenon. Finally, P can be defined as,
\begin{equation}
    P=\Delta G + \llbracket E \rrbracket - \kappa \sigma.
    \label{drivpress}
\end{equation}

\begin{figure}[!htb]
	\centering
	\includegraphics[width=13cm]{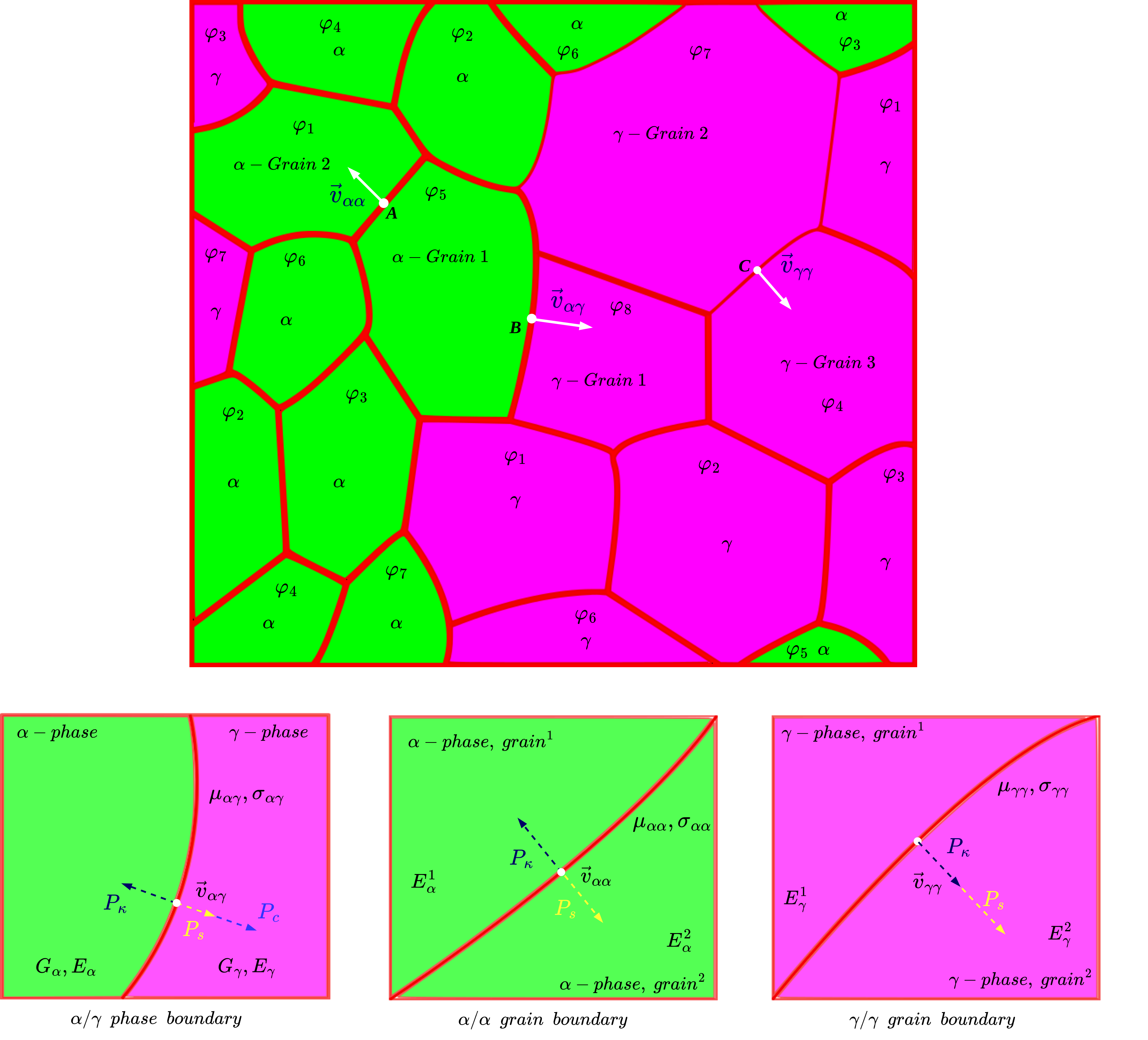}
	\captionsetup{justification=centering}
	\caption{A two-phase polycrystal with grains of phases $\alpha$ and $\gamma$ illustrating the velocity field at different types of interfaces}
	\label{classbiphpoly}
\end{figure}

The $\Delta G$ component acts only across the phase interfaces while vanishing across the grain interfaces of similar phases. Also, the sense and value of interface mobility, and interface energy could be different depending on the type of interface (i.e., $\alpha/\gamma$ phase interface, $\alpha/\alpha$ grain interface, and the $\gamma/\gamma$ grain interface). Thus, the velocity field, $\bm{v}$ needs to be dissected to be able to accommodate various driving pressure contributions relevant to specific interfaces. If we consider a classic two-phase polycrystal as illustrated in fig. \ref{classbiphpoly} with phases $\alpha$ and $\gamma$, $\bm{v}$ can be rewritten in the following form through interface characteristic functions:
\begin{equation}
\bm{v} = \chi_{\alpha \gamma}\bm{v}_{\alpha \gamma} + \chi_{\alpha \alpha}\bm{v}_{\alpha \alpha} + \chi_{\gamma \gamma}\bm{v}_{\gamma \gamma},
\label{kindiss}
\end{equation}where $\chi_{\alpha \gamma}$ is a characteristic function of the phase boundaries between the $\alpha$ and the $\gamma$ grains, $\chi_{\alpha \alpha}$ characterizes the grain boundaries between two $\alpha$ grains and likewise $\chi_{\gamma \gamma}$ for the grain boundaries between two $\gamma$ grains. 
\par
Hence, taking into account interface specific properties and driving pressures from the phase interfaces as well as the grain interfaces of both the parent and the product phase, we can formulate a generalized kinetic framework:
\begin{equation}
\begin{gathered}
\bm{v} = \chi_{\alpha \gamma}\mu_{\alpha \gamma}\left(\Delta G_{\alpha \gamma} - \kappa \sigma_{\alpha \gamma} + \llbracket E \rrbracket_{\alpha \gamma}\right)\bm{n} + \chi_{\alpha \alpha}\mu_{\alpha \alpha}\left(-\kappa \sigma_{\alpha \alpha}+\llbracket E \rrbracket_{\alpha \alpha}\right)\bm{n} \\
 + \chi_{\gamma \gamma}\mu_{\gamma \gamma}\left(-\kappa \sigma_{\gamma \gamma} + \llbracket E \rrbracket_{\gamma \gamma}\right)\bm{n}.
\end{gathered}
\label{totkin}
\end{equation}

Now, if we prescribe the above velocity field into eq.\eqref{lstranseqn}, and if we consider $\mathcal{S}=\{\alpha\gamma, \alpha\alpha, \gamma\gamma\}$, we get:
\begin{equation}
\begin{gathered}
\frac{\partial \varphi_i}{\partial t} + \left[\chi_{\alpha \gamma}\mu_{\alpha \gamma}\Delta G_{\alpha \gamma} + \sum_{l\in \mathcal{S}}\chi_{l}\mu_{l}\llbracket E \rrbracket_{l} \right]\bm{n}_i\cdot\bm{\nabla}\varphi_i - \left[\sum_{l\in \mathcal{S}}\chi_{l}\mu_{l}\sigma_{l}\right]\kappa_i \bm{n}_i\cdot\bm{\nabla}\varphi_i=0.
\end{gathered}
\label{LSeqnVpres}
\end{equation}

By verifying the metric property of a signed distance function, $\Vert \bm{\nabla}\varphi_i \Vert=1$ all along the simulation, we can write:

\begin{equation*}
\bm{n}_i=-\frac{\bm{\nabla}\varphi_i}{\Vert\bm{\nabla}\varphi_i\Vert}=-\bm{\nabla}\varphi_i \implies \kappa_i = \bm{\nabla}\cdot\bm{n}_i=-\Delta \varphi_i.
\end{equation*}

We can then rewrite eq.\eqref{LSeqnVpres} in a convective-diffusive form to be resolved for interface migration: 
\begin{equation}
\begin{gathered}
\frac{\partial \varphi_i}{\partial t} + \left[\bm{v_{\Delta G}}+\bm{v_{\llbracket E \rrbracket}}\right]_i\cdot\bm{\nabla}\varphi_i - \left[\sum_{l\in \mathcal{S}}\chi_{l}\mu_{l}\sigma_{l}\right]\Delta\varphi_i = 0 \qquad \forall i \in \{1,2,...,N_{LS}\},
\label{lsCDeqn}
\end{gathered}
\end{equation}where $\bm{v_{\Delta G}}=\chi_{\alpha \gamma}\mu_{\alpha \gamma}\Delta G_{\alpha \gamma} \bm{n}$, and $\bm{v_{\llbracket E \rrbracket}}=\left[\sum_{l\in \mathcal{S}}\chi_{l}\mu_{l}\llbracket E \rrbracket_{l} \right]\bm{n}$.

The resolution of the above equations must be followed by a reinitialization procedure to restore the metric properties of the LS functions at each time step. This ensures $\varphi_i$ regularity, thus preserving good conditioning of the LS transport equation. Conserving signed Euclidean distance functions also ensure that the above LS transport eqs.\eqref{lsCDeqn} remain true in their convective-diffusive form (which allows for avoiding an exact calculation of the curvature term). In addition, by keeping $\varphi_i$ a distance function, some parts of the global level-set resolution such as the remeshing algorithms can be properly based on the notion of Euclidean distance to the interface. In the context of this work, a recent reinitialization strategy  \cite{shakoor2015efficient} that involves a fast, direct calculation based on an optimized brute force algorithm is adopted.
\subsubsection*{Better description for $\bm{v_{\Delta G}}$:}
In the context of a polycrystal with multiple junctions, for the convective part of eq.\eqref{lsCDeqn}, the above description of $\bm{v_{\Delta G}}$ (and also $\bm{v_{\llbracket E \rrbracket}}$) is not sufficient if we seek to avoid discontinuous velocity fields or kinematic incompatibilities at the multiple junctions. For that, it is more efficient to work with a common velocity field for all the $N_{LS}$ level-set functions and the velocity field needs to be as regular as possible around the multiple junctions. In their work dedicated to the simulation of recrystallization in single-phase polycrystals, Bernacki et al \cite{Bernacki2009Rex} proposed the following formulation for $\bm{v_{\llbracket E \rrbracket}}$:
\begin{equation}
    \bm{v_{\llbracket E \rrbracket}}(\bm{x},t)=\sum_{i=1}^{N_{LS}}\sum_{\substack{j=1 \\ j \neq i}}^{N_{LS}}\chi_{G_i}(\bm{x},t)\mu_{ij}\exp{\left(-\beta|\varphi_j|\right)}\llbracket E \rrbracket_{ij}(\bm{x},t)(-\bm{n}_j),
    \label{vstoredBernacki}
\end{equation}where $\chi_{G_i}$ is the characteristic function of the grain $G_i$, $\mu_{ij}$ is the interface mobility between the neighbouring grains $i$ and $j$, the exponential term is a continuous decreasing function varying from $1$ to $0$ on either side of the interface and has the function of smoothening the velocity field across the interface, $\beta$ is a positive parameter that controls the degree of smoothness, $\llbracket E \rrbracket_{ij}(\bm{x},t)=\mathcal{E}_j(\bm{x},t)-\mathcal{E}_i(\bm{x},t)$ is the jump in stored energy of two neighbouring grains $i$ and $j$ where $\mathcal{E}_j(\bm{x},t)$ and $\mathcal{E}_i(\bm{x},t)$ can be the average stored energies of the grains $i$ and $j$ respectively \cite{Bernacki2009Rex} or more local approximations \cite{Ilin2018}, and $\bm{n}_j$ is the outward unit normal to the neighbouring grain $j$.

\begin{figure}[!htb]
	\centering
	\captionsetup{justification=centering}
	\subfloat[$\vec{v}_{\llbracket E \rrbracket}$ component]{\includegraphics[width=13cm]{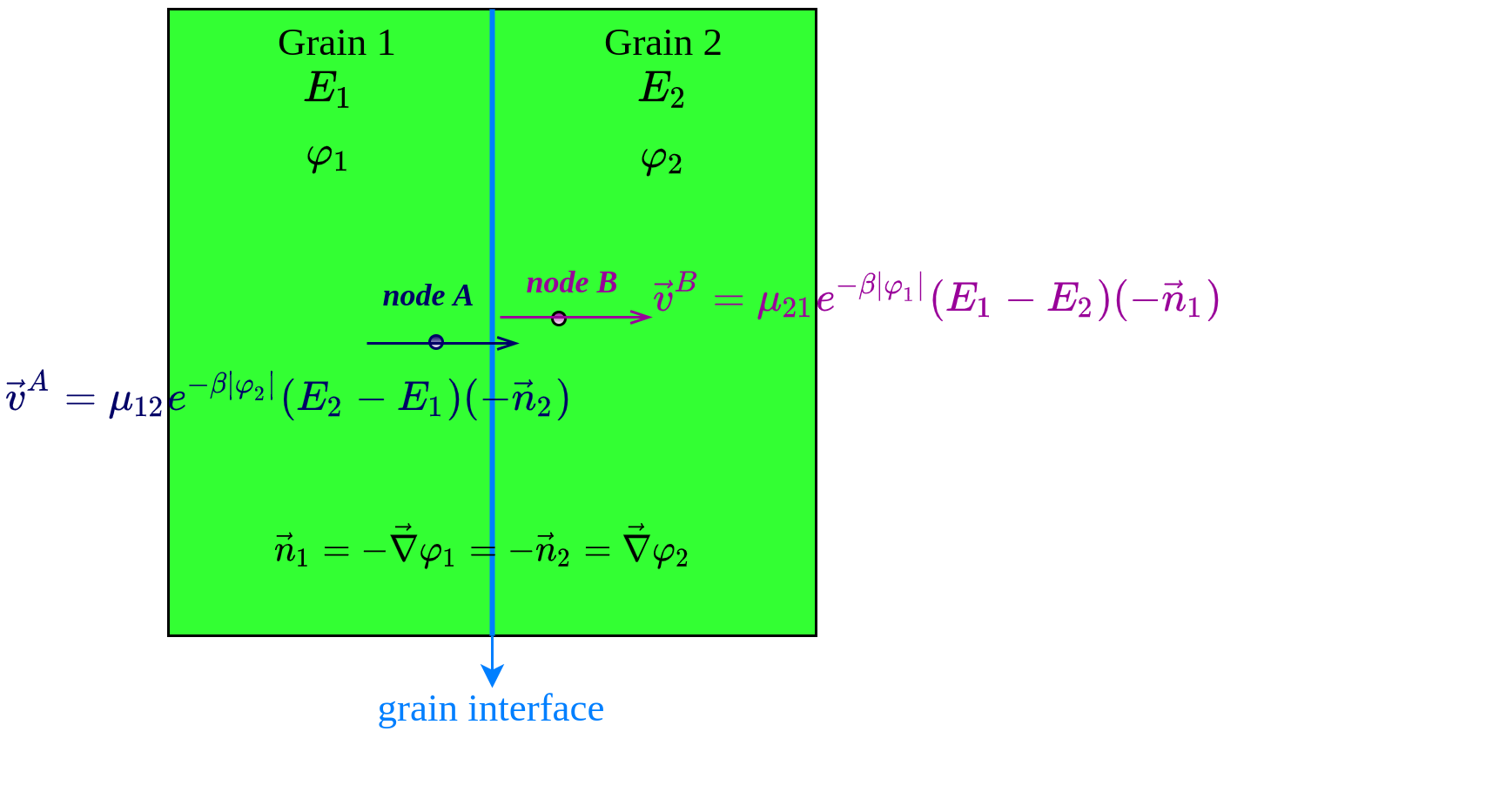}\label{vstoredsignflip}}\\
	\subfloat[$\vec{v}_{\Delta G}$ component]{\includegraphics[width=13cm]{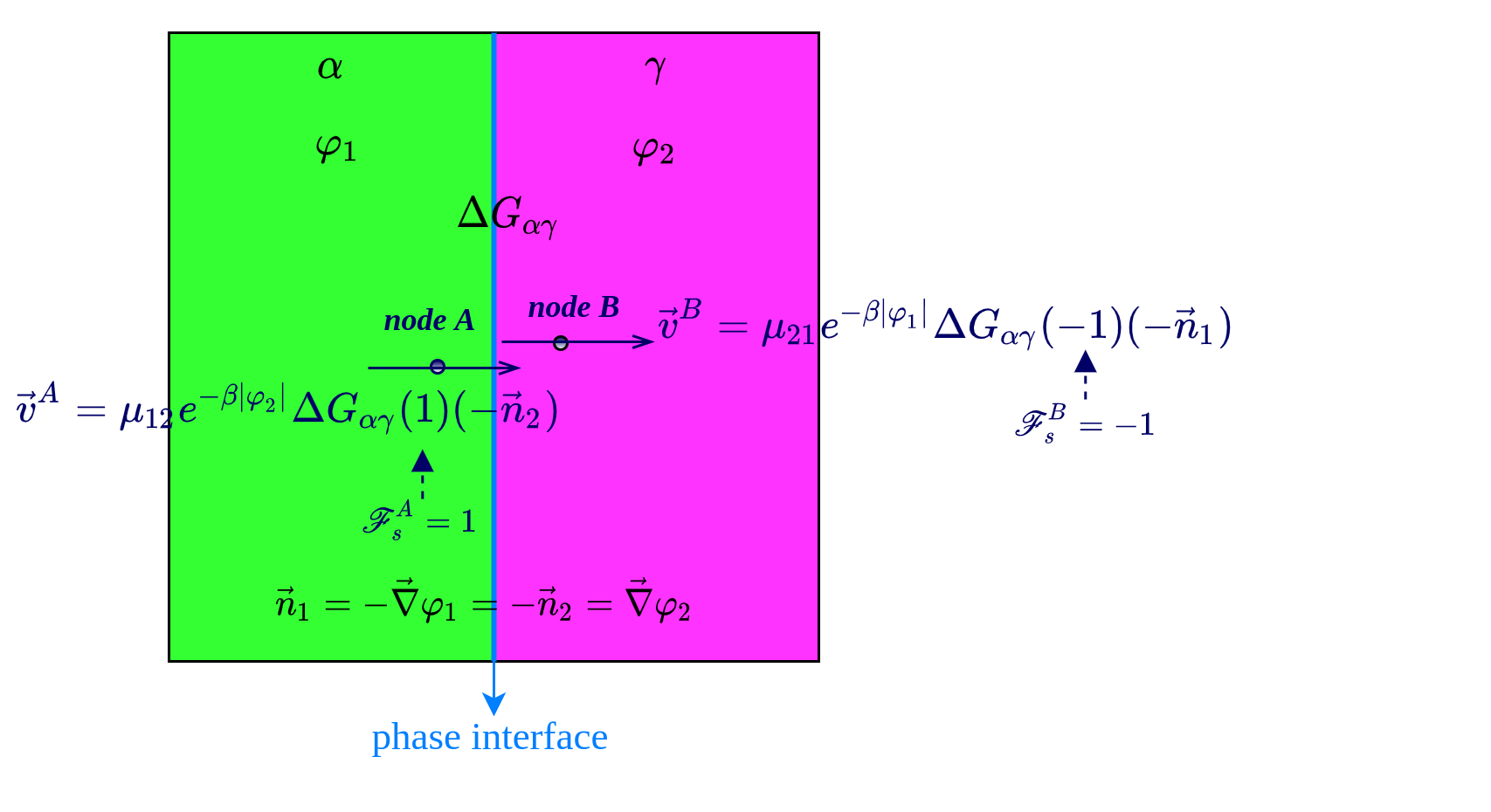}\label{vdGsignflip}}
	\caption{Consistent orientation of velocity vectors on nodes A and B, close to either side of the migrating interface}
	\label{velorient}
\end{figure}

So for $\bm{v_{\Delta G}}$ component, we take inspiration from the above equation and propose an analogous formulation, albeit with a couple of additional functions:
\begin{equation}
    \bm{v_{\Delta G}}(\bm{x},t)=\sum_{i=1}^{N_{LS}}\sum_{\substack{j=1 \\ j \neq i}}^{N_{LS}}\chi_{G_i}\mu_{ij}\exp{\left(-\beta|\varphi_j|\right)}\chi_{\alpha\gamma}\Delta G_{\alpha\gamma}\mathscr{F}_s(-\bm{n}_j),
    \label{vdeltaG}
\end{equation}where $\chi_{\alpha\gamma}$, as seen earlier, helps to filter this component of velocity field only on the phase interfaces. In eq.\eqref{vstoredBernacki}, the jump in stored energies $\llbracket E \rrbracket_{ij}$ ensures that the velocity vectors are oriented in a consistent direction on the nodes close to both the sides of the interface, thanks to a flip in sign as shown in fig. \ref{vstoredsignflip}. However in eq.\eqref{vdeltaG}, since $\Delta G_{\alpha\gamma}$ already gives a measure of the Gibbs free energy difference on the phase interface, there is no natural flip in sign. Hence $\mathscr{F}_s$ is used as a sense function that ensures that the velocity vectors of this component on the nodes close to either side of the phase interface are oriented consistently as observed in fig. \ref{vdGsignflip}. $\mathscr{F}_s$ in the context of austenite decomposition ($\gamma \rightarrow \alpha$) is defined as follows:
\begin{equation}
    \mathscr{F}_s(\bm{x},t)=\chi_\alpha(\bm{x},t)-\chi_\gamma(\bm{x},t)=2\chi_\alpha(\bm{x},t)-1,
    \label{sensefunc}
\end{equation}where $\chi_\alpha(\bm{x},t)$ and $\chi_\gamma(\bm{x},t)$ are the characteristic functions of $\alpha$ and $\gamma$ phase respectively.

\subsubsection*{Description for $\Delta G_{\alpha\gamma}$:}\label{DGdescsection}

The last ingredient missing to completely prescribe the above kinetics is the change in Gibbs free energy between the two phases. $\Delta G_{\alpha\gamma}$ is typically dependent on the local composition of the solutes, temperature, and the pressure. In many works, the description for $\Delta G_{\alpha\gamma}$ has been
established by thermodynamic evaluations based on Calphad data \cite{steinbach2007calphad} or ThermoCalc software \cite{thermocalc}. For certain sharp interface descriptive models, the diffusion in the product phase is assumed to be instantaneous and so $\Delta G_{\alpha\gamma}$ is simply assumed to be proportional to the deviation in concentration at the interface in the parent phase ($C_{\gamma, eq}$) from the equilibrium concentration in this phase ($C_{\gamma, \gamma \alpha}$) \cite{mecozzi2011quantitative, liu2018review}:
\begin{equation*}
    \Delta G_{\alpha\gamma}=\Upsilon\left(C_{\gamma, eq}-C_{\gamma, \gamma \alpha}\right),
    \label{deltaGold}
\end{equation*}where $\Upsilon$ is a proportionality factor that could be temperature dependent and is derived
from thermodynamic databases.

\begin{figure}[!htb]
	\centering
	\captionsetup{justification=centering}
	\subfloat[$\alpha/(\alpha+\gamma) \ \ boundary$]{\includegraphics[width=8cm]{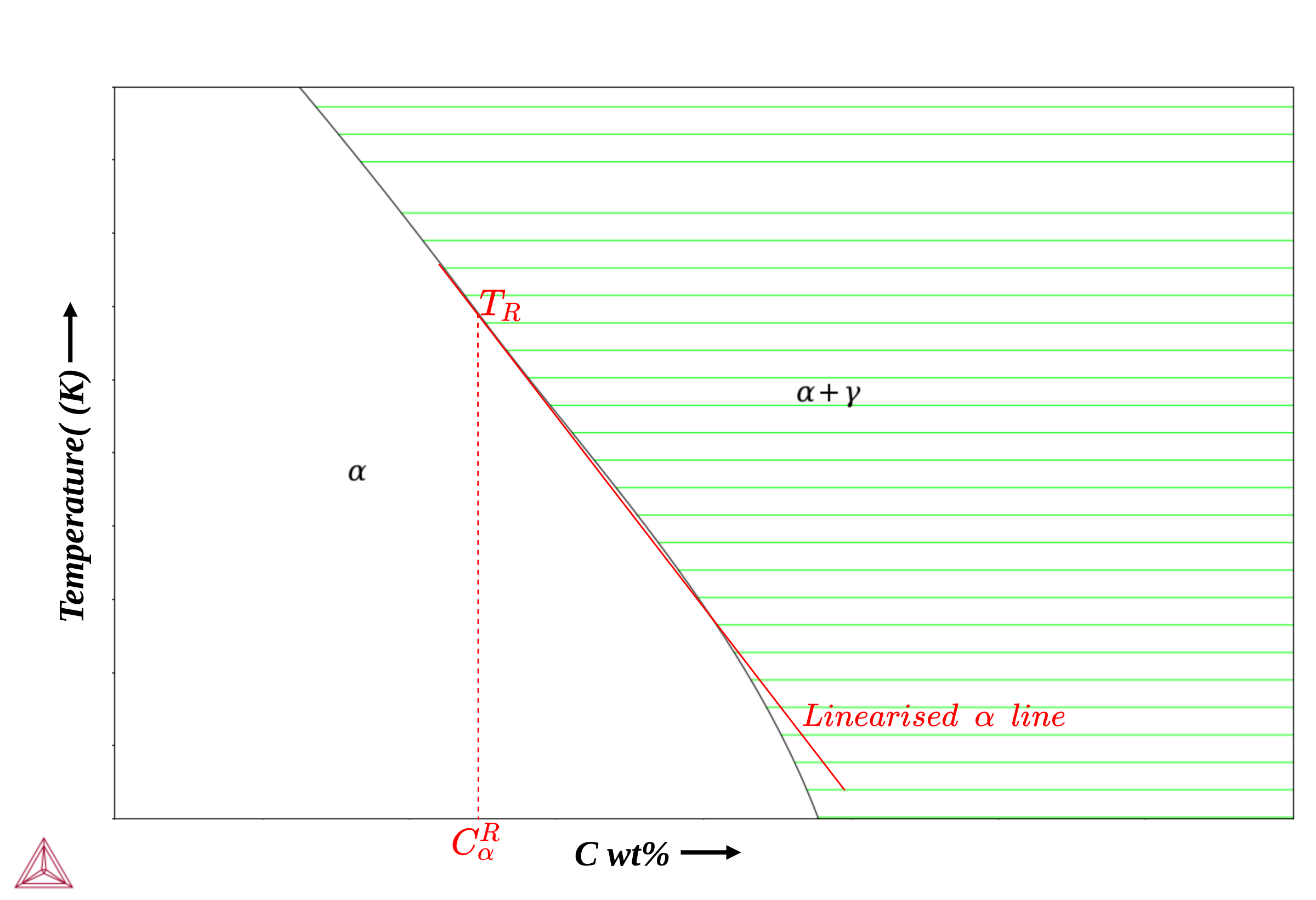}\label{lpdalpha}}
	\subfloat[$\gamma/(\alpha+\gamma) \ \ boundary$]{\includegraphics[width=8cm]{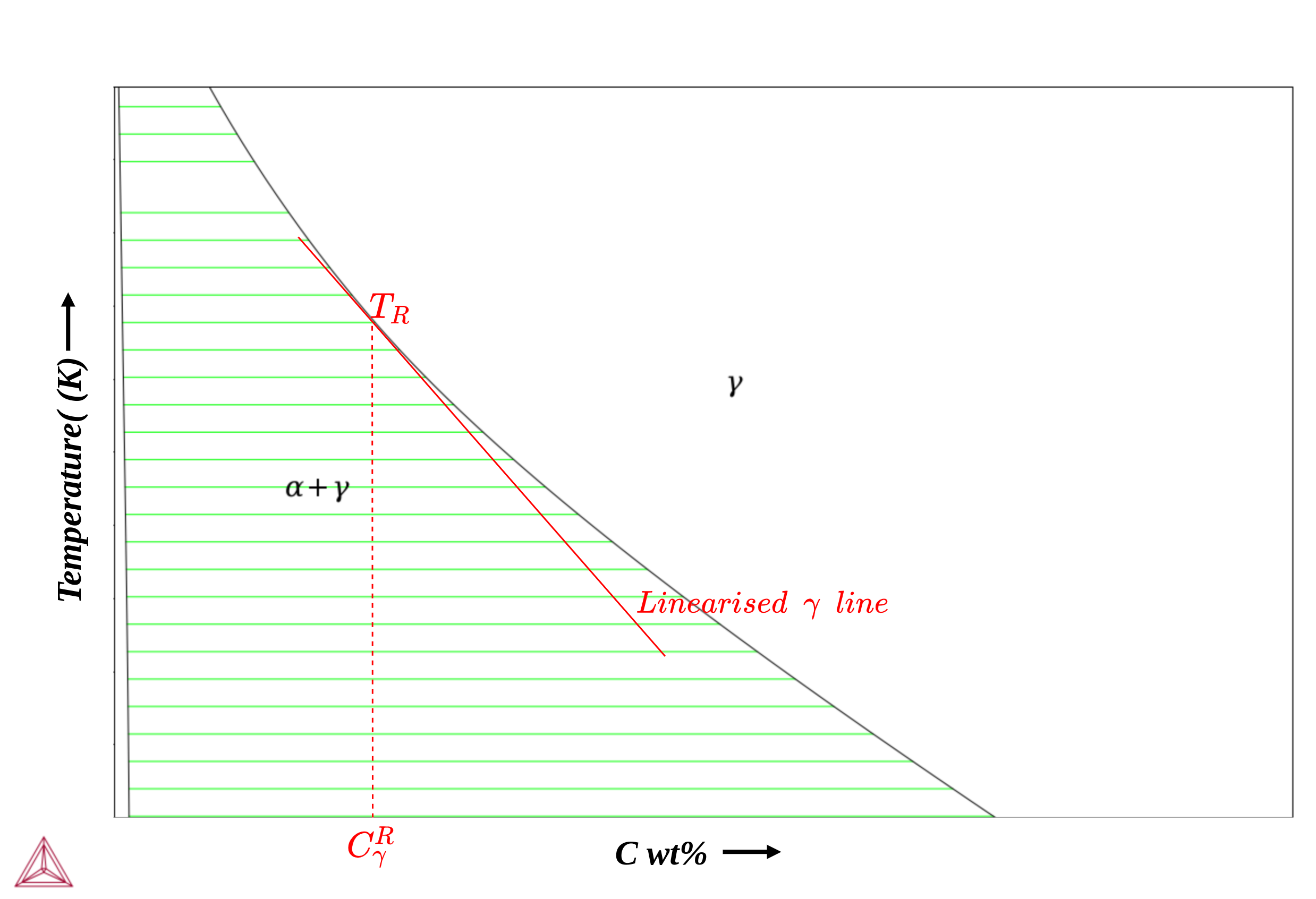}\label{lpdgamma}}
	\caption{Linearized phase diagrams with phase boundaries linearized at $T^R$ }
	\label{thermoCdemolinph}
\end{figure}

In the current work, $\Delta G_{\alpha\gamma}$ is described based on a local linearization of the phase diagram as seen in the works of Mecozzi et al \cite{Mecozzi2007PhaseFM}. $\Delta G$ is basically assumed to be proportional to a small undercooling ($\Delta T = T^{eq}-T$). At low undercooling with the assumptions that the actual temperature $T$ is close to the equilibrium temperature $T^{eq}$ (corresponding to a local composition of $C_\alpha$ and $C_\gamma$), the variations of enthalpy ($\Delta H$), and the entropy ($\Delta S$) with temperature could be considered to be negligible ($\Delta S^{eq} \approx \Delta S$, $\Delta H^{eq} \approx \Delta H$) \cite{hillert2007phase, Pariser2001}. $\Delta G$ is thus given by:  
\begin{equation}
\Delta G(T,C) =  \frac{\Delta H^{eq}}{T^{eq}}(T^{eq}-T)=\Delta S^{eq} \Delta T.
\label{DeltaGprop}
\end{equation}
Linearizing at a reference temperature ($T^R$), and assuming only carbon element partitions, we can write:
\begin{equation}
    \begin{gathered}
    T^{eq}_\alpha=T^R + m^R_\alpha \left(C_\alpha - C_\alpha^R\right), \\
    T^{eq}_\gamma=T^R + m^R_\gamma \left(C_\gamma - C_\gamma^R\right),
    \end{gathered}
    \label{linatR}
\end{equation}where $m_\alpha^R$ and $m_\gamma^R$ are the slopes of the boundary lines of the $\alpha$ and $\gamma$ phases respectively, linearized at $T^R$. $C_\alpha^R$ and $C_\gamma^R$ are the equilibrium carbon concentrations at $T^R$ of ferrite and austenite respectively. These are deduced by thermodynamic evaluations using ThermoCalc software \cite{thermocalc} as shown in fig. \ref{thermoCdemolinph}. 

The undercooling is expressed as:
\begin{equation}
    \Delta T=\frac{T_\alpha^{eq}+T_\gamma^{eq}}{2}-T
    \label{deltaTdef}
\end{equation}
So, if we substitute the two eqs. in \eqref{linatR} into eq.\eqref{deltaTdef}, we can then write $\Delta T$ as:
\begin{equation}
    \Delta T=T^R +  0.5m^R_\alpha\left(C_\alpha - C_\alpha^{R}\right)+  0.5m^R_\gamma\left(C_\gamma - C_\gamma^{R}\right) -T.
    \label{DeltaTdesc}
\end{equation}
From eq.\eqref{DeltaGprop} and eq.\eqref{DeltaTdesc}, $\Delta G_{\alpha\gamma}$ component is then expressed as a function of the local concentrations and temperature as follows:
\begin{equation}
\Delta G_{\alpha\gamma} (T, C_\alpha, C_\gamma) = \Delta S\left[(T^R - T) + 0.5m^R_\alpha\left(C_\alpha - C_\alpha^{R}\right) + 0.5m^R_\gamma\left(C_\gamma - C_\gamma^{R}\right) \right].
\label{fenergy1}
\end{equation}
With the help of eqs.\eqref{cmix} and \eqref{keq}, the above description could be further expressed as a function of the total concentration variable, $C$, for each configuration of the phase-field function, $\phi (\bm{x},t)$:
\begin{equation}
\begin{gathered}
\Delta G = \Delta S\left[T^R - T + 0.5m^R_\gamma\left(\frac{C}{1 + \phi(k-1)} - C_\gamma^{R}\right) + 0.5m^R_\alpha\left(\frac{k	C}{1 + \phi(k-1)} - C_\alpha^{R}\right)\right]
\end{gathered}
\label{fenergy2}
\end{equation}

Based on the same linearization, the equilibrium carbon concentrations of each phase at temperature T can be estimated as follows:
\begin{equation}
C_i^{eq}=C_i^R+\frac{T-T^R}{m_i^R}, \quad with \ \ i =\{\alpha, \gamma\}.
\label{ceqestim}
\end{equation}
Using eqs.\eqref{ceqestim}, the equilibrium partitioning ratio ($k$) can be expressed at each temperature $T$ as:
\begin{equation}
k(T) = \frac{C_\alpha^R+\frac{T-T^R}{m_\alpha^R}}{C_\gamma^R+\frac{T-T^R}{m_\gamma^R}}
\label{keqfinal}.
\end{equation}
The presence of a jump in stored energy $\llbracket E\rrbracket$ due to a plastic deformation will not be considered in the illustrative test cases of this article but will be discussed in a forthcoming publication. So, in the following, $\bm{v}_{\llbracket E \rrbracket}$ is neglected. 

\subsection{Additional numerical considerations in the context of polycrystals:}
To simulate DSSPT in the context of polycrystals with $N_{LS}$ global level-set functions ($\varphi_i$), along with the above numerical formalism, we need to consider some supplementary fields and particular numerical treatments to support certain aspects of the simulation. 

\subsubsection*{Computation of phase-field variable, $\phi$:}
In order to compute the phase-field function using the hyper tangent relation in eq.\eqref{tanhf} at each time, we need a signed distance LS function that represents all the zones of the product ferrite phase ($\varphi_{\alpha-zone}$) in the overall domain.  In order to facilitate the computation of such a function, we need to use the characteristic function of the $\alpha$ phase ($\chi_\alpha(\bm{x},t)$). This phase characteristic function, $\chi_\alpha$, is updated at each time step after the resolution of eq.\eqref{lsCDeqn}. Then, $\varphi_{\alpha-zone}$ is obtained through the reinitialization of the $ \mathscr{F}_s(\bm{x},t)\varphi_{max}(\bm{x},t)=\left(2\chi_\alpha(\bm{x},t)-1\right)\varphi_{max}(\bm{x},t)$ function (with $\varphi_{max}(\bm{x},t)=\max\limits_{i=1,...,N_{LS}} \varphi_i(\bm{x},t)$) through an imposed $2\epsilon$ thickness around its 0-isovalue. A similar result can be obtained by a direct reinitialization of the $\mathscr{F}_s(\bm{x},t)$ function but the interest to multiply it by the $\varphi_{max}(\bm{x},t)$ function is to increase the 0-isocontour precision before the reinitialization.  The hyper tangent relation eq.\eqref{tanhf} is then applied to $\varphi_{\alpha-zone}$ to compute the phase-field function.
\begin{figure}[!htb]
	\centering
	\captionsetup{justification=centering}
	\subfloat[$\alpha$-phase characteristic field with the interfaces highlighted in black for visualization]{\includegraphics[width=7.5cm]{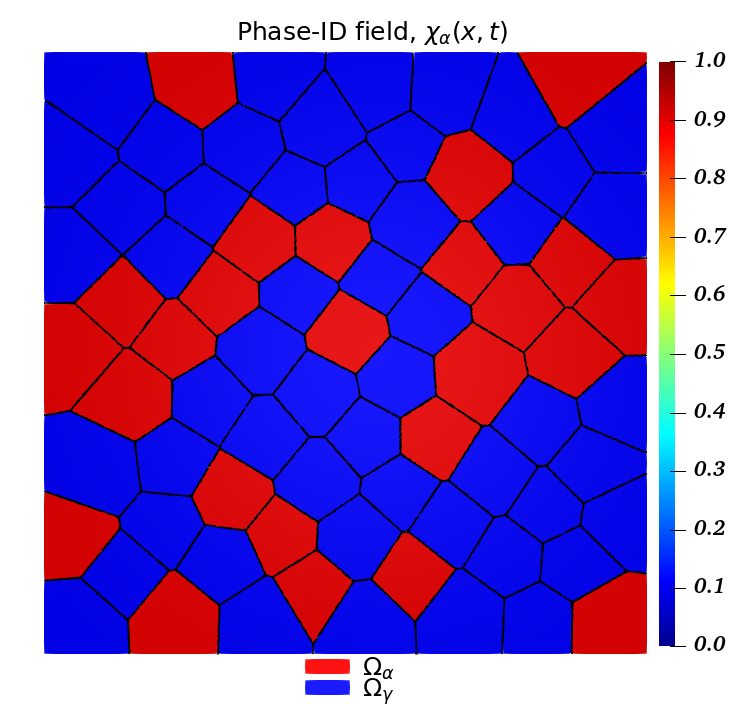}\label{phaseidrep}}\\
	\subfloat[Ferrite zone signed distance function]{\includegraphics[width=7.5cm]{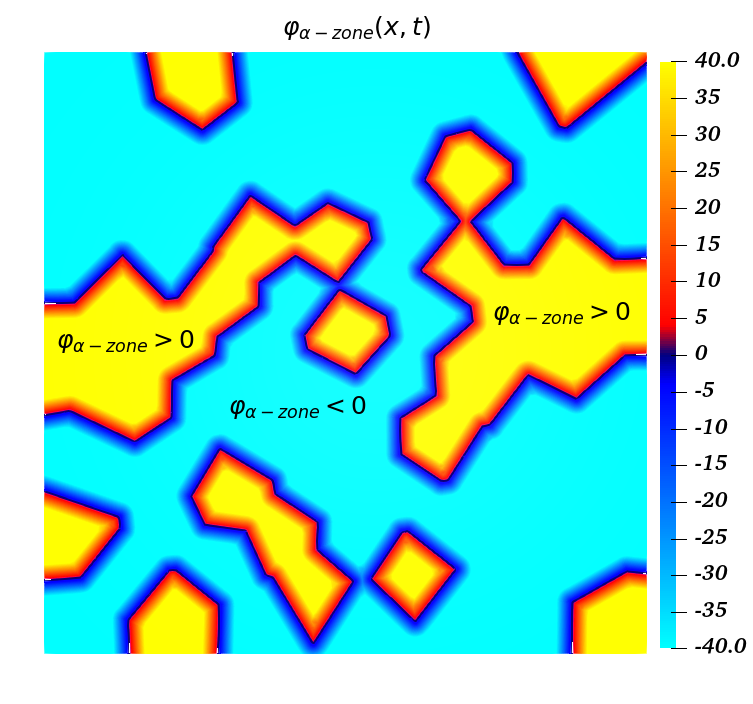}\label{AZlsf}}
    \subfloat[Phase-field function]{\includegraphics[width=7.5cm]{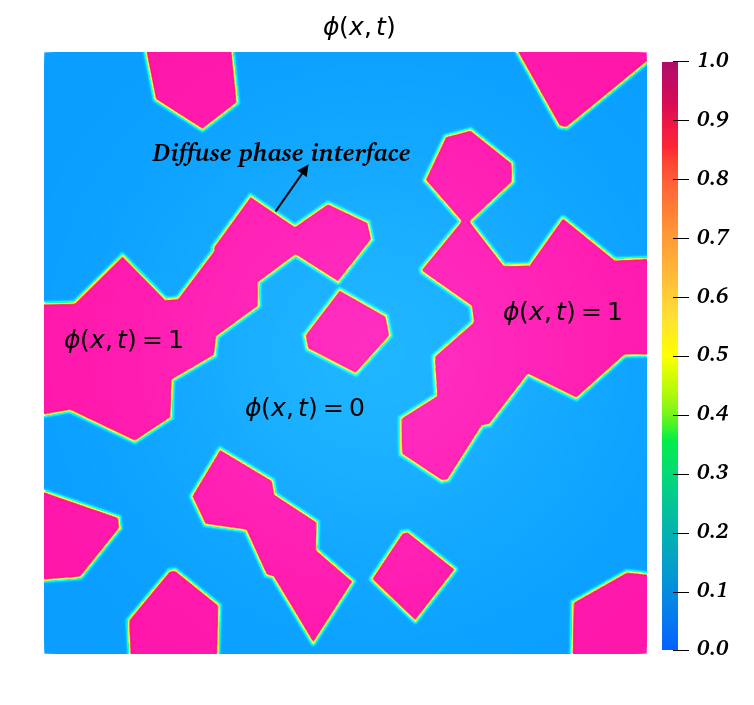}\label{pfrep}}
	\caption{Illustrations for computing phase-field function in a two-phase polycrystal}
	\label{addfuncrep}
\end{figure}
Fig. \ref{addfuncrep} illustrates the methodology. 
\subsubsection*{Interface characteristic functions:}
\begin{figure}[!htb]
	\centering
	\captionsetup{justification=centering}
	\subfloat[$\alpha/\gamma$ phase interfaces]{\includegraphics[width=5.5cm]{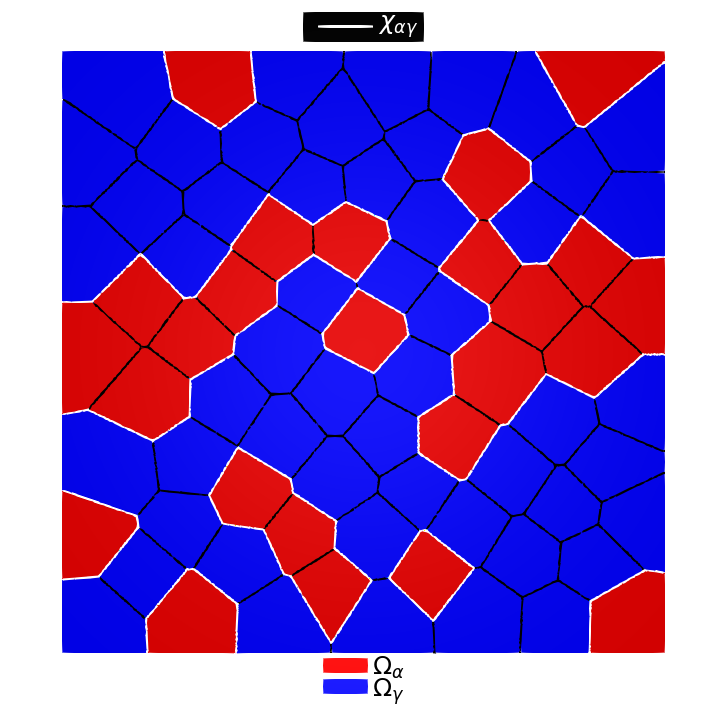}\label{charagrep}}
	\subfloat[$\alpha/\alpha$ grain interfaces]{\includegraphics[width=5.5cm]{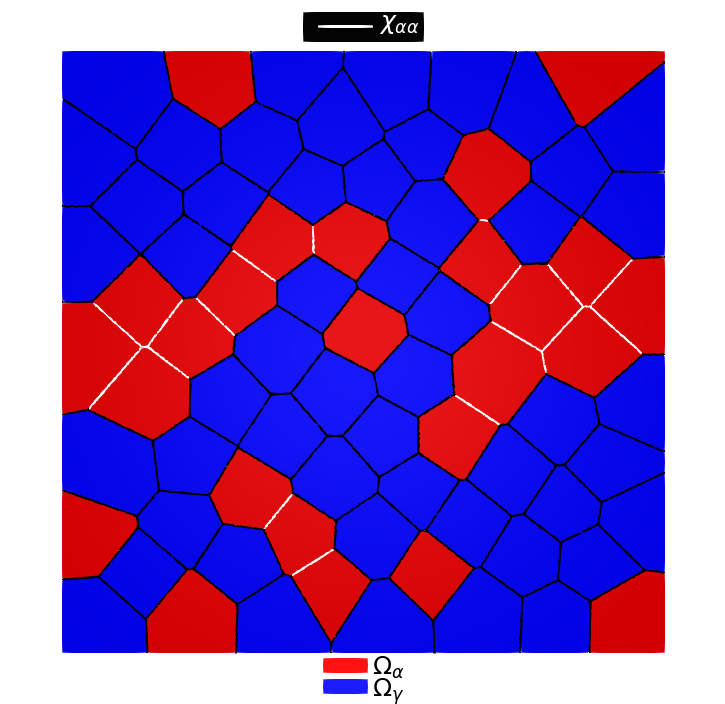}\label{charaarep}}
    \subfloat[$\gamma/\gamma$ grain interfaces]{\includegraphics[width=5.5cm]{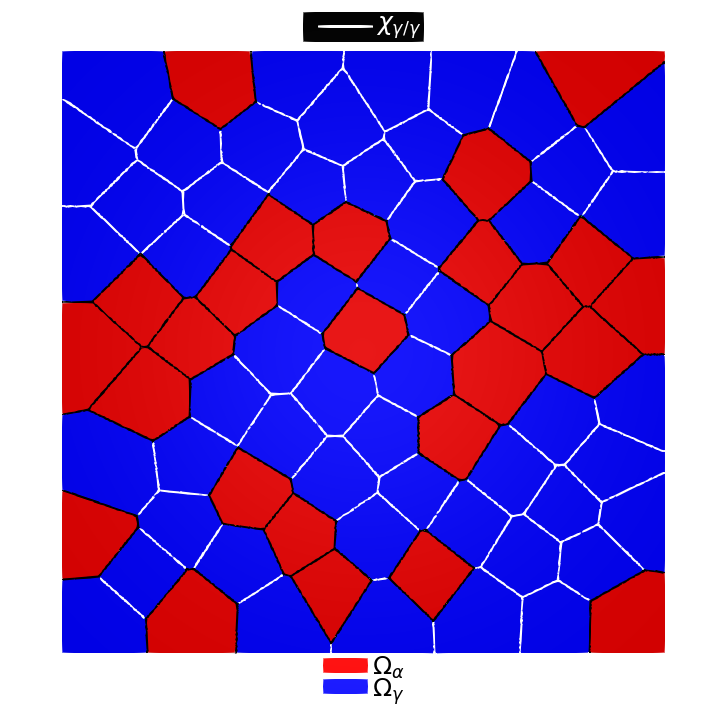}\label{charggrep}}
	\caption{Illustrations of various interface characteristic functions in a two-phase polycrystal. Respective interfaces are highlighted in white}
	\label{intcharfuncs}
\end{figure}
The interface characteristic functions are computed at each instant as follows:
\begin{equation}
    \begin{gathered}
    \chi_{\alpha\gamma}(\bm{x},t)=
    \begin{cases}
    1 & if \ \ |\varphi_{\alpha-zone}(\bm{x},t)|<\epsilon \\
    0 & otherwise
    \end{cases}\\
     \chi_{\alpha\alpha}(\bm{x},t)=
     \begin{cases}
    \left[1-\chi_{\alpha\gamma}(\bm{x},t)\right]\chi_\alpha(\bm{x},t) & if \ \ \varphi_{max}(\bm{x},t)< \epsilon \\
    0 & otherwise
    \end{cases}\\
     \chi_{\gamma\gamma}(\bm{x},t)=
     \begin{cases}
    1-\chi_{\alpha\gamma}(\bm{x},t)-\chi_{\alpha\alpha}(\bm{x},t) & if \ \ \varphi_{max}(\bm{x},t)< \epsilon \\
    0 & otherwise
    \end{cases}
    \end{gathered}.
\end{equation}Fig. \ref{intcharfuncs} shows an illustration of these functions.

\subsubsection*{Numerical treatment at the multiple junctions and reinitialization:}
Following the LS transport resolution, due to the presence of multiple junctions, in order to remove any kinematic incompatibilities at the multiple junctions such as vacuum or overlapping regions, a particular numerical treatment according to \cite{Merriman1994MotionOM} is performed to modify the LS functions:
\begin{equation}
    \hat{\varphi_i}=\frac{1}{2}\left(\varphi_i-\max_{j\neq i}\varphi_j\right) \quad \forall i\in \{1,...,N_{LS}\}.
\label{mjtreat}
\end{equation}
Following this multiple junctions treatment, $\varphi_i(\bm{x},t)$ are reinitialized in the $2\epsilon$-narrow band around their 0-isovalues at each time step. The term $\epsilon$ is taken to be equal to at least $2$ times the $\eta$ value to ensure that $\varphi_i$, $\varphi_{max}$, and hence $\varphi_{\alpha-zone}$ are all regular and well defined far enough from the corresponding interfaces such that $\phi$ is properly computed for the considered $\eta$ parameter value.

The simulations presented in the next section were carried out with unstructured triangular meshes, a P1 interpolation, and using an implicit backward Euler time scheme for the time discretization. Each system linked to eq.\eqref{diffWF} and the weak formulation of eqs.\eqref{lsCDeqn} is assembled using typical P1 FE elements with a Streamline Upwind Petrov-Galerkin (SUPG) stabilization for the convective terms \cite{Brooks1982}. The boundary conditions (BCs) are classical null-von Neumann BCs applied to all the LS functions and carbon concentration. Each plane of the boundary domain can be seen as a symmetric plane for the LS functions and the carbon field. 
\section{Results and discussion}
\label{resdisc}
The following hypotheses have been imposed for the illustration cases considered here to simulate DSSPT:
\begin{itemize}
	\item The character of the phase transformation kinetics are assumed to be of \textbf{mixed-mode} \cite{sietsma2004concise} with both interface and diffusion controlled modes. So, the solute concentration at the interface doesn't attain the equilibrium concentration right away, and the diffusion in the bulk of the phase is not instantaneous. 
	
	\item \textbf{Para-equilibrium} \cite{hillert2004definitions} conditions are assumed. In other words, the partitioning of any substitutional solute elements such as manganese is neglected since the diffusion of such elements is generally several orders slower than that of interstitial elements. So only interstitial elements such as carbon are assumed to be redistributed and contribute to the $\Delta G$ driving pressure.
	
	\item The interface mobility, $\mu$, and the interface energy, $\sigma$ are both assumed to be isotropic for now. The interface mobility is given a temperature dependence through an Arrhenius type law \cite{arhenius}:
	\begin{equation*}
	\mu = \mu_0\exp{\left(-\frac{Q_\mu}{RT}\right)},
    \end{equation*}
    where, $Q_\mu$ is the activation energy for grain boundary migration taken as \SI{140}{\kilo\joule\per\mole} \cite{krielaart1998kinetics}, $R$ is the universal gas constant, and $\mu_0$ is the pre-exponential factor taken as \SI{2e17}{\micro\meter\tothe{4}\per\joule\per\second} for low cooling rates or as \SI{6e17}{\micro\meter\tothe{4}\per\joule\per\second} for high cooling rates \cite{Mecozzi2007PhaseFM}. Diffusivities of the two phases are also assumed temperature dependent using a similar Arrhenius type law. The diffusivity pre-factors are taken from \cite{Mecozzi2007PhaseFM} as well. 
\end{itemize}

\subsection{Pseudo-1D case with planar interface:}
\begin{figure}[!htb]
	\centering
	\includegraphics[width=12cm]{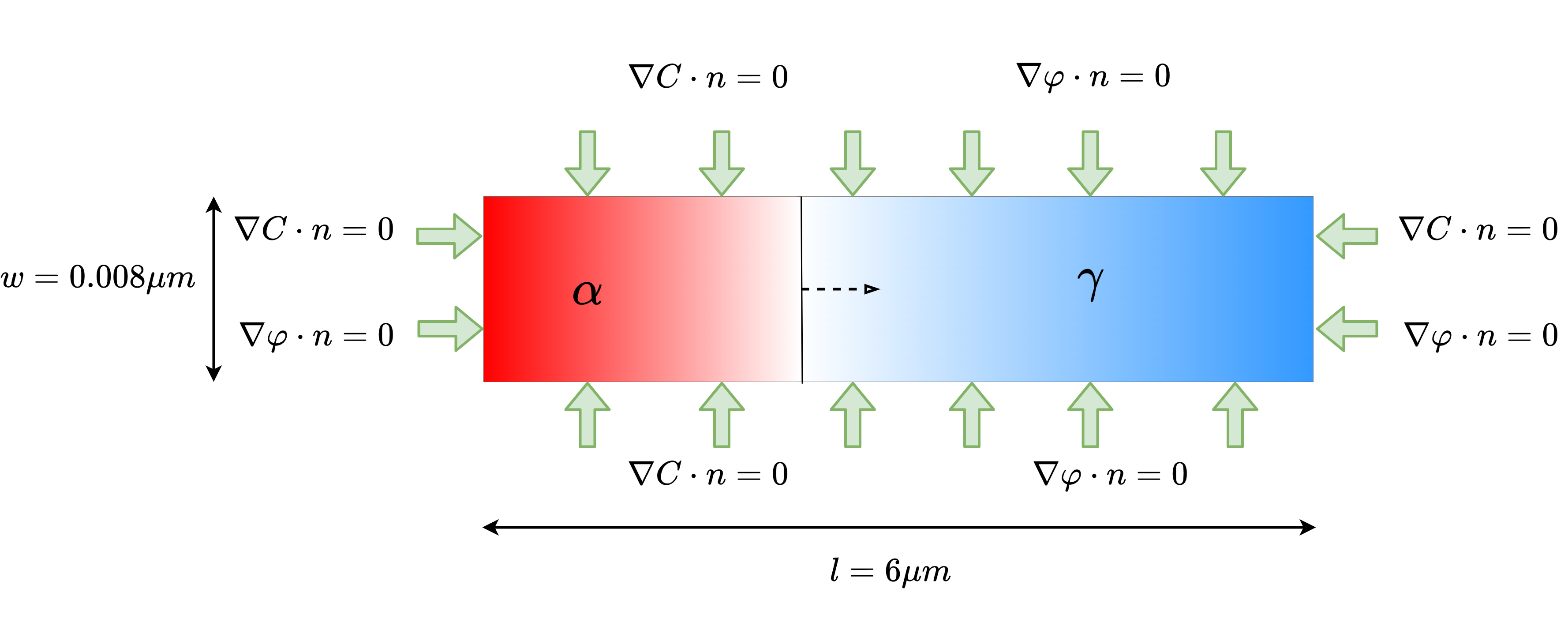}
	\captionsetup{justification=centering}
	\caption{Pseudo-1D case ($w<<l$) with a planar interface between a ferrite (red) and an austenite (blue) phase}
	\label{1Dtestcase}
\end{figure}
As a first case, we consider a slender 2D domain (Pseudo-1D since $w << l$) with a planar interface between one austenite grain and one ferrite grain as shown in fig. \ref{1Dtestcase}. A simple material with a composition of $Fe-C$ \SI{0.02}{\wtpercent} is assumed. The initial condition is assumed to be at a temperature, $T^i=$ \SI{1173}{\kelvin}, with corresponding initial concentrations of $C_\alpha^i=$ \SI{0.0014022}{\wtpercent} and $C_\gamma^i=$ \SI{0.024575}{\wtpercent} (extracted from ThermoCalc). The $\alpha / \gamma$ phase interface is initially imposed to be at $\Gamma^i=$ \SI{1.1838}{\micro\meter} from the left boundary. A reference temperature of $T^R=$ \SI{1160}{\kelvin} is taken and the necessary thermodynamic data (summarized in table \ref{1DTrdata}) are extracted using ThermoCalc. The final state is imposed to be at a temperature, $T^f=$ \SI{1140}{\kelvin}, and the corresponding equilibrium data are summarized in table \ref{1DTfdata}. The final steady state interface position is expected to be at, $\Gamma^{eq}=$ \SI{5.11296}{\micro\meter}. No capillarity effects are considered for the planar interface (null curvature). The thickness of the diffuse phase interface for this case is taken as $\eta=$ \SI{0.5}{\micro\meter}. The FE mesh is static with a mesh size, $h$ equals to \SI{0.8}{\nano\meter}. The time step is fixed to \SI{0.2}{\milli\second}.
\begin{table}[!htb]
\centering
 \begin{tabular}{ c | c | c | c | c | c } 
\specialrule{.2em}{.1em}{.1em}  
 $T^R$ (\SI{}{\kelvin}) & $C_{\alpha}^R$ (\SI{}{\wtpercent})& $C_{\gamma}^R$ (\SI{}{\wtpercent}) & $\Delta S$ (\SI{}{\joule\per\kelvin\per\micro\meter\cubed})& $m_\alpha^R$ (\SI{}{\kelvin\per\wtpercent}) & $m_\gamma^R$ (\SI{}{\kelvin\per\wtpercent})\\ [0.5ex] 
 \hline
 $1160$ & $0.0029083$ & $0.054289$ & $2.8481175\times10^{-13}$ & $-8746.564$ & $-416.959$ \\ [1ex] 
\specialrule{.2em}{.1em}{.1em}
 \end{tabular}
 \captionsetup{justification=centering}
 \caption{ThermoCalc data extracted at $T^R$ for $Fe-C$ \SI{0.02}{\wtpercent}}
 \label{1DTrdata}
\end{table}

\begin{table}[!htb]
\centering
 \begin{tabular}{ c | c | c | c } 
\specialrule{.2em}{.1em}{.1em}  
 $T^f$ (\SI{}{\kelvin})& $C_{\alpha}^{eq}$ (\SI{}{\wtpercent})& $C_{\gamma}^{eq}$ (\SI{}{\wtpercent})& Ferrite fraction, $f_\alpha^{eq}$ \\ [0.5ex] 
 \hline
 1140 & 0.0051473 & 0.10593 & 0.85216 \\ [1ex] 
\specialrule{.2em}{.1em}{.1em}
 \end{tabular}
 \captionsetup{justification=centering}
\caption{Expected steady state at $T^f$}
 \label{1DTfdata}
\end{table}

We consider three different scenarios of cooling: (i) instantaneous cooling from \SI{1173}{\kelvin} to \SI{1140}{\kelvin}, thus giving isothermal phase transformation, (ii) rapid cooling rate of \SI{10}{\kelvin\per\second}, and (iii) gradual cooling rate of \SI{3}{\kelvin\per\second}. 
\begin{figure}[!htb]
	\centering
	\includegraphics[width=15cm]{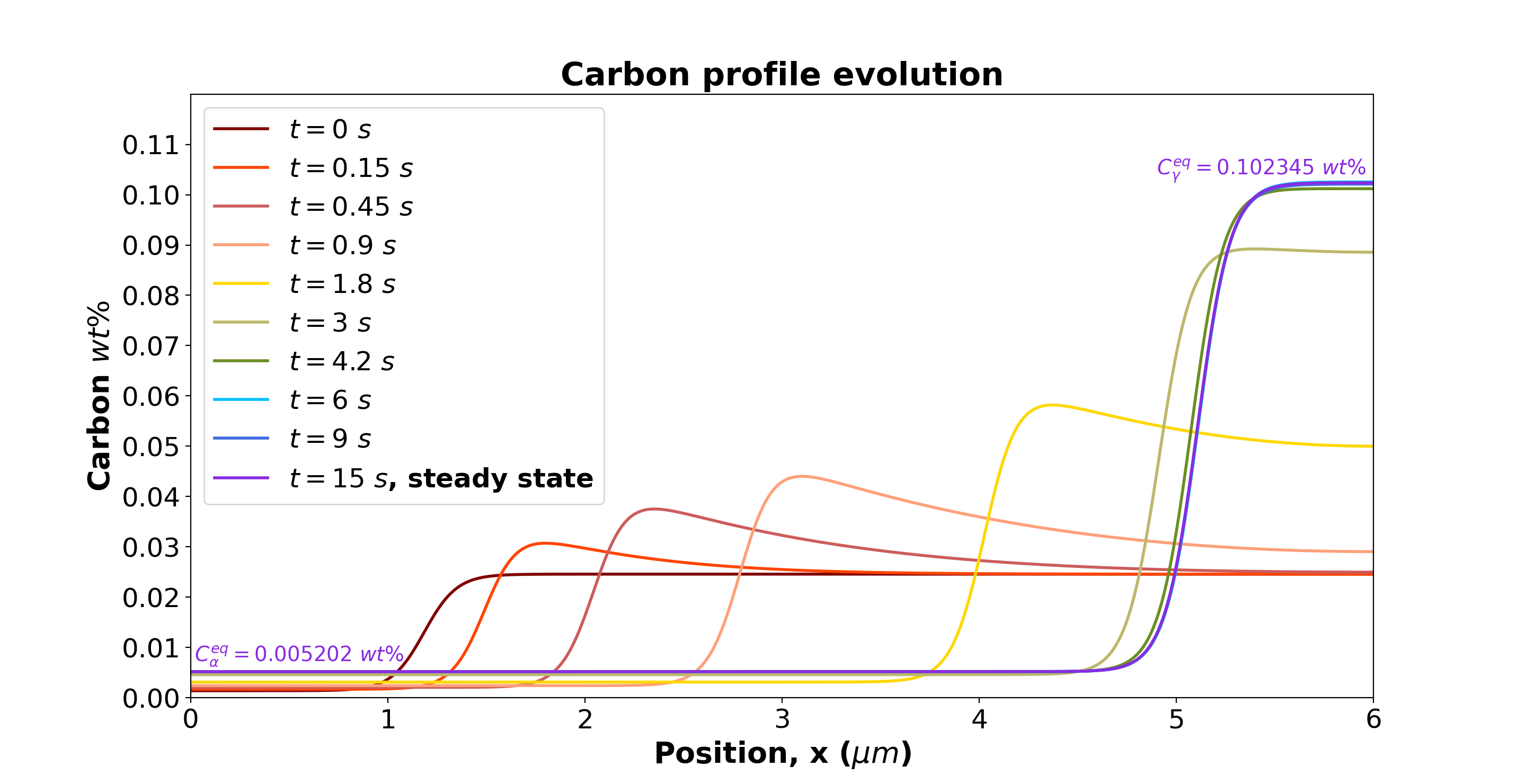}
	\captionsetup{justification=centering}
	\caption{Evolution of carbon profiles at different instants till the steady state for the case with instantaneous cooling}
	\label{allcprof}
\end{figure}

\begin{figure}[!htb]
	\centering
	\captionsetup{justification=centering}
	\subfloat[Phase interface evolution]{\includegraphics[width=7.5cm]{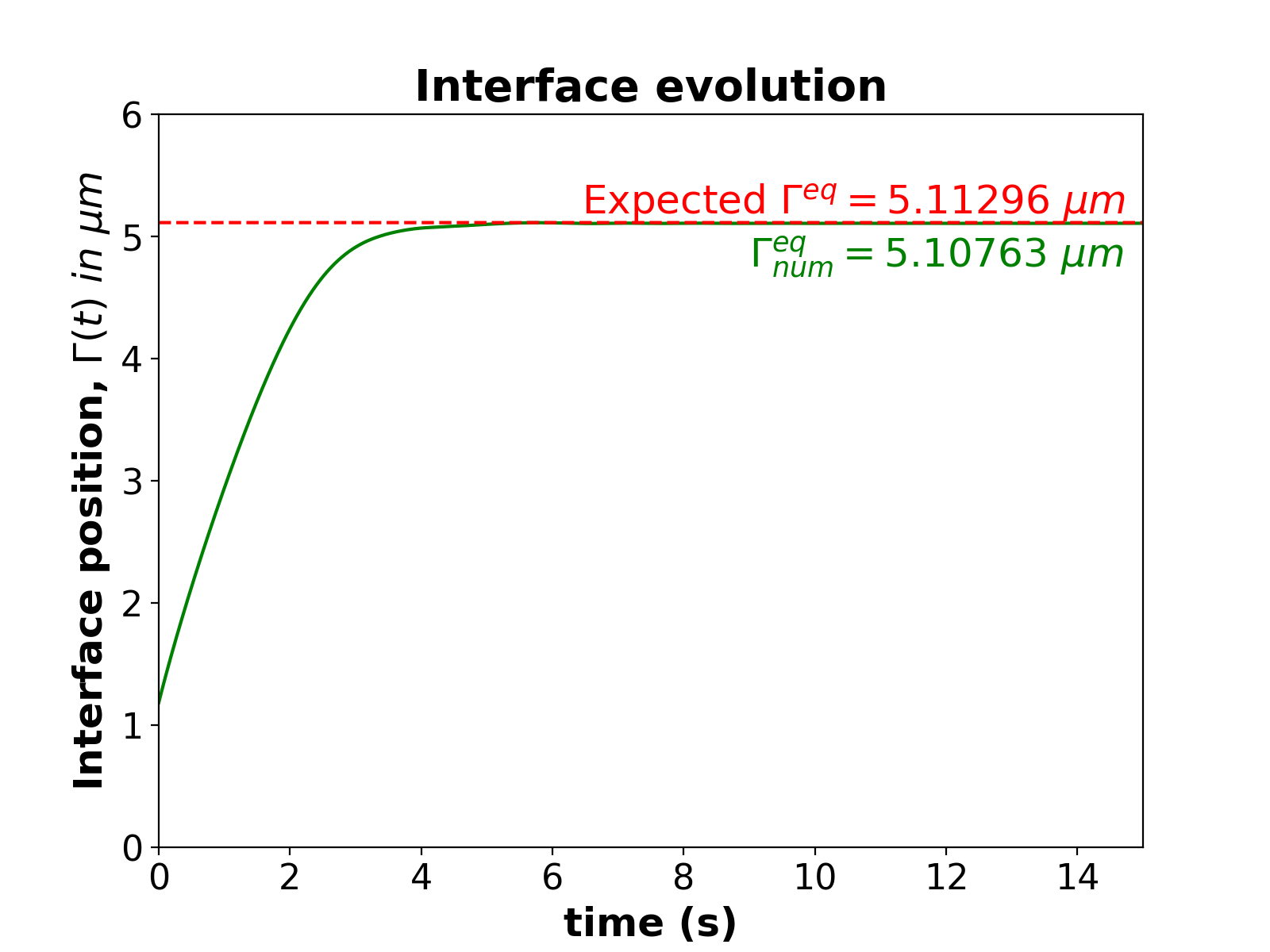}\label{1Dintevol}}
	\subfloat[Solute mass variation]{\includegraphics[width=7.5cm]{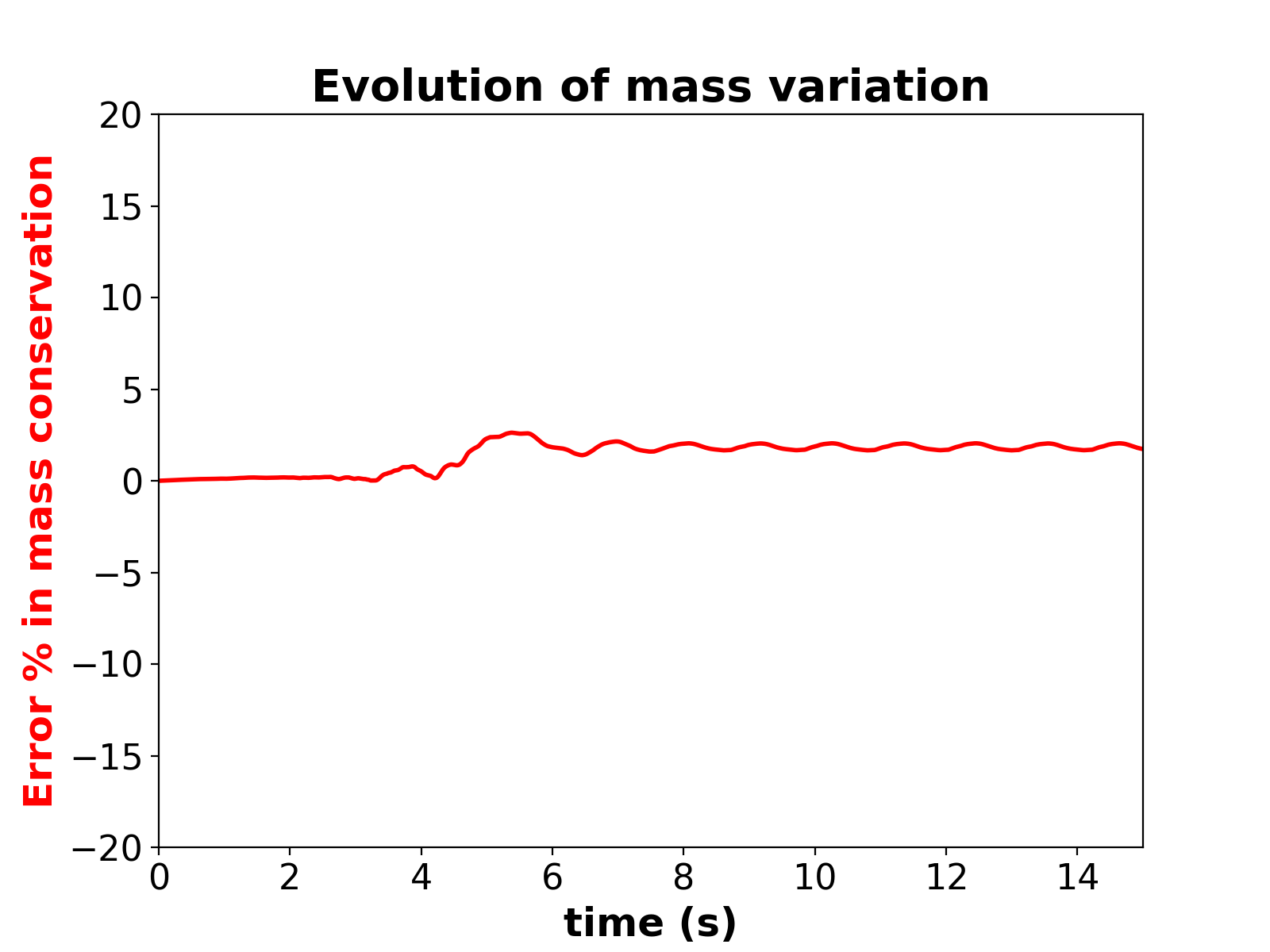}\label{1Dmcerror}}
	\caption{Interface migration to a steady state position, and the variation of solute mass in the domain for the instantaneous cooling case} 
	\label{1DintermceEvol}
\end{figure}

Fig. \ref{allcprof} illustrates the evolution of carbon concentration at different times for the instantaneous cooling case. As the interface starts to migrate, we can observe the development of peaks in the profiles close to the interface on the austenite side as the ferrite phase rejects carbon into the austenite phase. This indicates the expected solute enrichment in austenite during the transformation as carbon is generally more soluble in austenite than ferrite. Carbon profiles on the ferrite side are mostly plain since the diffusivity of carbon in ferrite is higher, and hence diffusion is faster compared to that in austenite. The concentration in the austenite side continues to increase until the steady state between the two phases with the corresponding equilibrium concentrations. As the steady state is reached, plain carbon profiles are obtained in both phases. At steady state, the simulated equilibrium concentrations are found to be: \textcolor{red}{$C_\alpha^{eq, \ num}=$ \SI{0.005202}{\wtpercent}} and \textcolor{red}{$C_\gamma^{eq, \ num}=$ \SI{0.10234}{\wtpercent}}. Fig \ref{1Dintevol} describes the interface evolution converging to its steady state position of \textcolor{red}{$\Gamma_{num}^{eq}=$ \SI{5.1076}{\micro\meter}} equivalent to an equilibrium ferrite fraction of \textcolor{red}{$f_ \alpha^{eq, num}=$ 0.85127}. The steady state values obtained are in close agreement with the expected state tabulated in table \ref{1DTfdata}. Minor differences in numerically predicted concentrations from the expected values stem from the linearization of the phase diagram. For the cases with continuous cooling, the steady state predictions could be further improved by considering multiple reference points ($T^{R_1}, T^{R_2}, ..., T^{R_n}$) properly spaced along the considered thermal path ($T^i$ to $T^f$) in the phase diagram and by extracting necessary data from ThermoCalc at multiple reference temperatures. Fig. \ref{1Dmcerror} quantifies the quality of mass conservation during the course of the simulation. The maximum variation is limited to $2.6223 \%$. This variation generally stems from the mesh quality, the choice of the time step, diffuse interface thickness ($\eta$) et cetera. Solute mass variation is also found to be more prominent as the interface gets closer to the domain boundaries. This is because of the nature of the boundary conditions imposed as opposed to the sense of the resulting solute flux from interface migration. 
\begin{figure}[!htb]
	\centering
	\includegraphics[width=10cm]{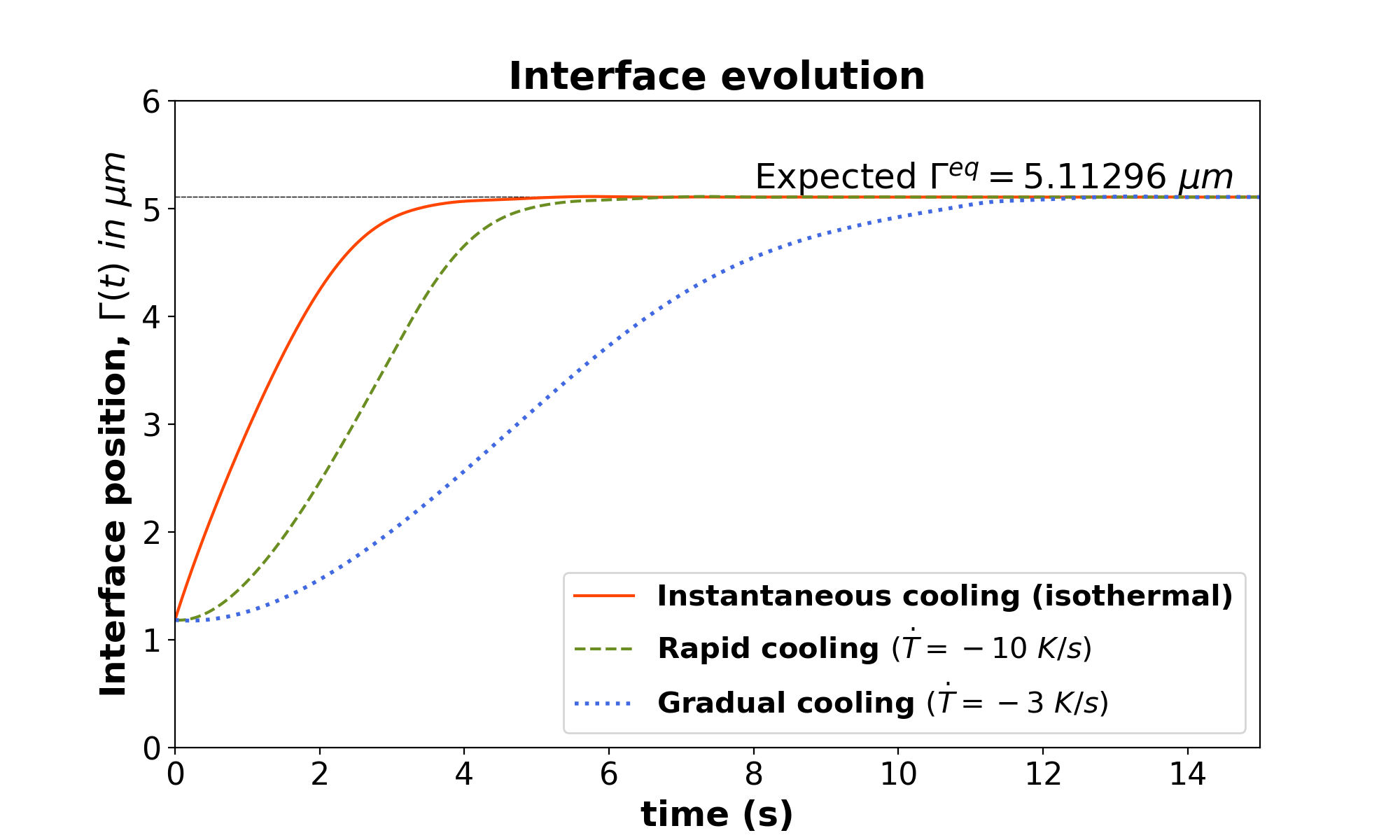}
	\captionsetup{justification=centering}
	\caption{Comparison of the kinetics of interface evolution for the different scenarios of cooling}
	\label{1DCoolingComp}
\end{figure}

Fig. \ref{1DCoolingComp} compares the kinetics of interface evolution for different cases of cooling. Clearly, the case with gradual cooling is slow to start as it steadily departs from the initial equilibrium state and is the slowest to reach equilibrium. On the other hand, the case with instantaneous cooling takes off immediately and swiftly reaches the steady state. All three cases yield similar steady states with differences only in their kinetics.  
\subsubsection*{Comparison with a semi-analytic 1D sharp interface model:}
\begin{figure}[!htb]
	\centering
	\includegraphics[width=12cm]{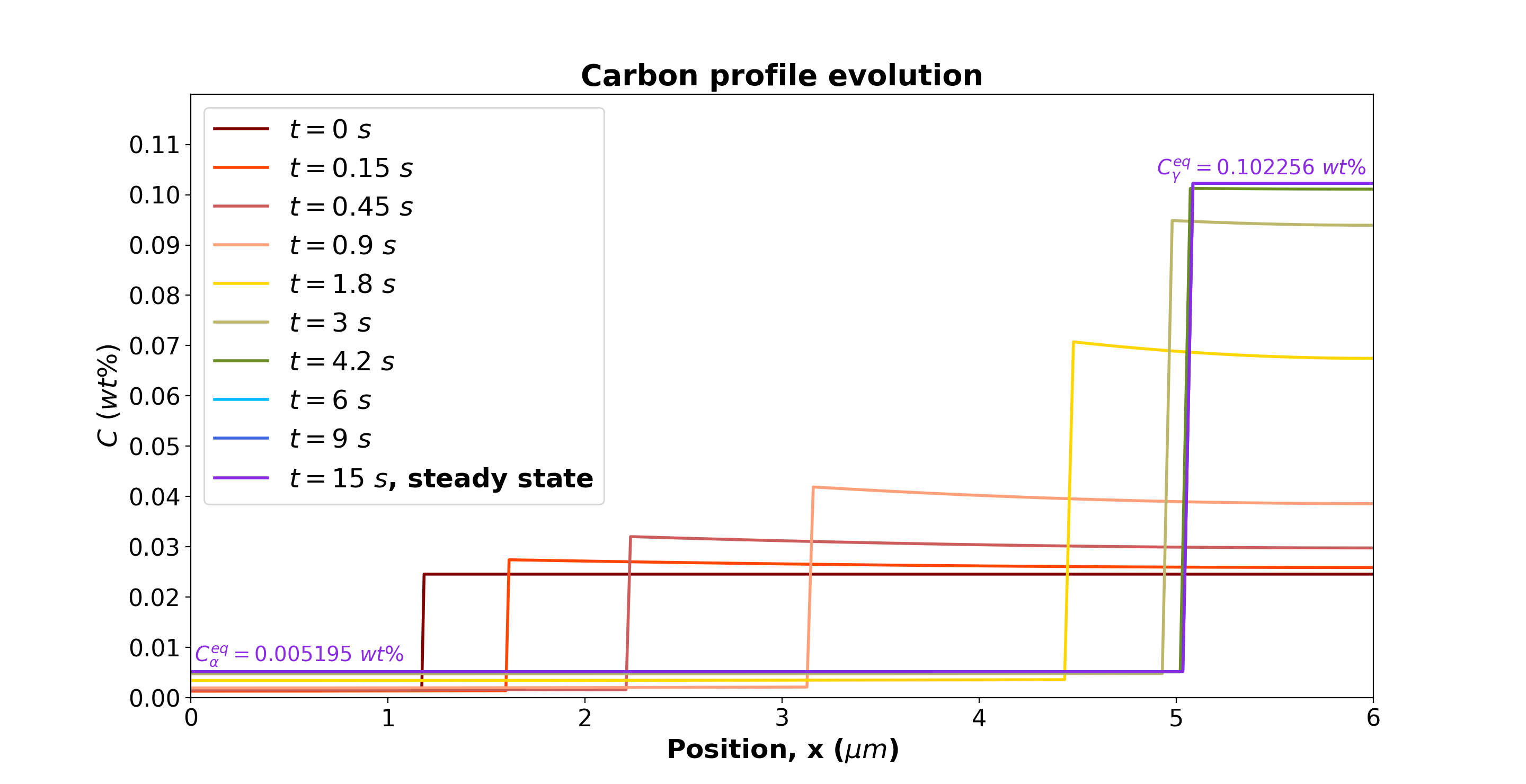}
	\captionsetup{justification=centering}
	\caption{Carbon profiles predicted by the semi-analytical model for the case with instantaneous cooling}
	\label{allcprofSA}
\end{figure}

\begin{figure}[!htb]
	\centering
	\includegraphics[width=8cm]{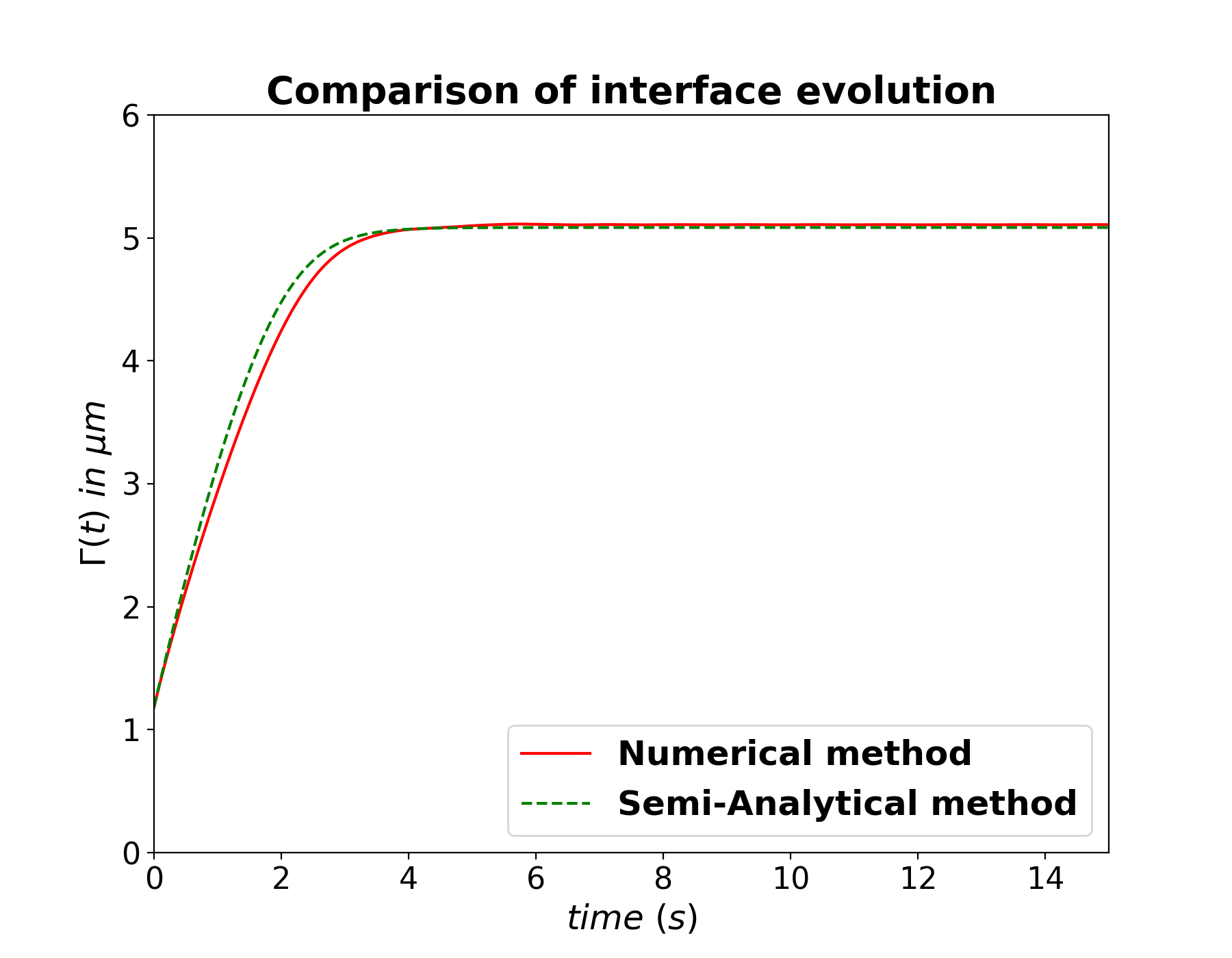}
	\captionsetup{justification=centering}
	\caption{Comparison of interface evolution predicted by the semi-analytical and the LS based numerical model}
	\label{1DintevolSAvNUM}
\end{figure}

\begin{figure}[!htb]
	\centering
	\captionsetup{justification=centering}
	\subfloat[$C_\alpha^{int}$ evolution]{\includegraphics[width=7.5cm]{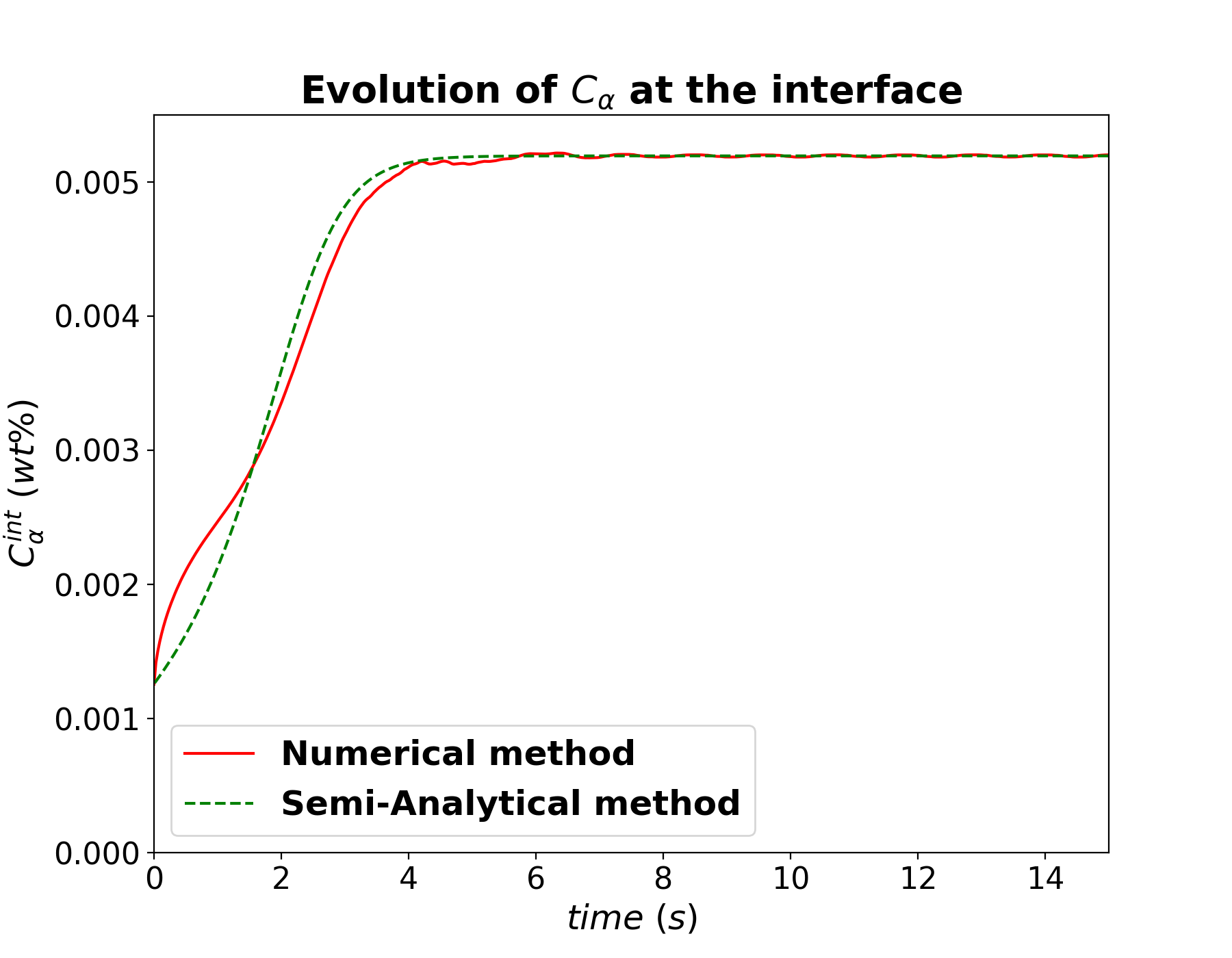}\label{CaintevolSAvNUM}}
	\subfloat[$C_\gamma^{int}$ evolution]{\includegraphics[width=7.5cm]{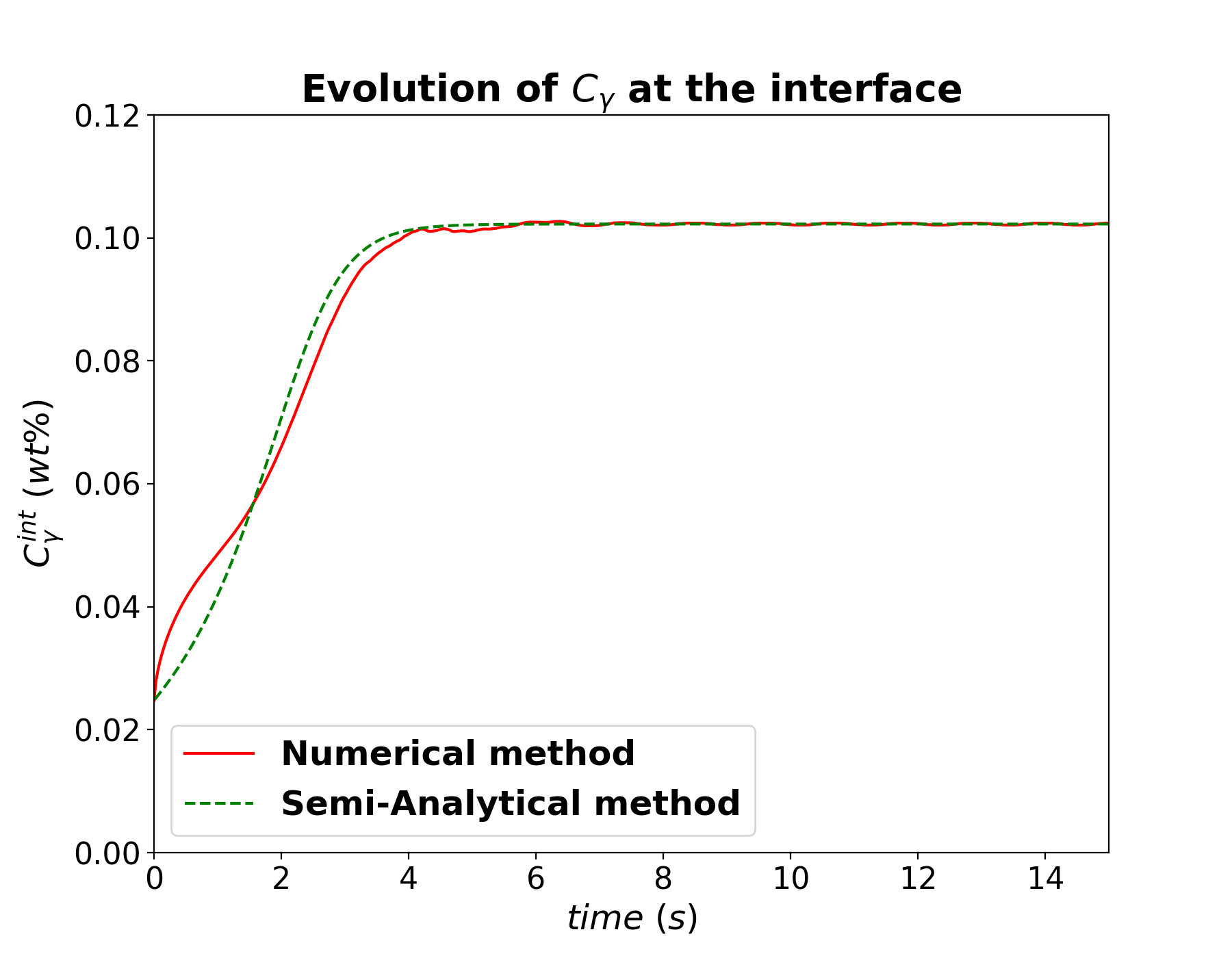}\label{CgintevolSAvNUM}}
	\caption{Comparison of the evolution of carbon concentration at the interface predicted by the semi-analytical and the numerical model}
	\label{CintSAvNUM}
\end{figure}

A 1D semi-analytic sharp interface model for mixed-mode phase transformation was implemented to be able to compare with the LS predictions. The reader is referred to the appendix - \ref{semianalytic} for more details on this semi-analytic formulation which is an extension of the model proposed in \cite{chen2011modeling}. Fig. \ref{allcprofSA} shows the evolution of carbon profiles along with the equilibrium concentrations predicted by the semi-analytical model for the instantaneous cooling case. The concentrations obtained correspond closely with that of the LS Simulations. Fig. \ref{1DintevolSAvNUM} compares the kinetics of interface evolution by the two methods. The kinetics and also the steady state interface position obtained are in good agreement between the two methods. Figs. \ref{CintSAvNUM} demonstrates that the interface concentrations predicted by the two methods are congruent which explains the good agreement in interface kinetics since both methods use the same description of the driving pressure.

It has been found that there is good accordance between the two methods for the non-isothermal scenarios also.  
\subsection{2D two-phase polycrystal case:}
\begin{figure}[!htb]
	\centering
	\includegraphics[width=9cm]{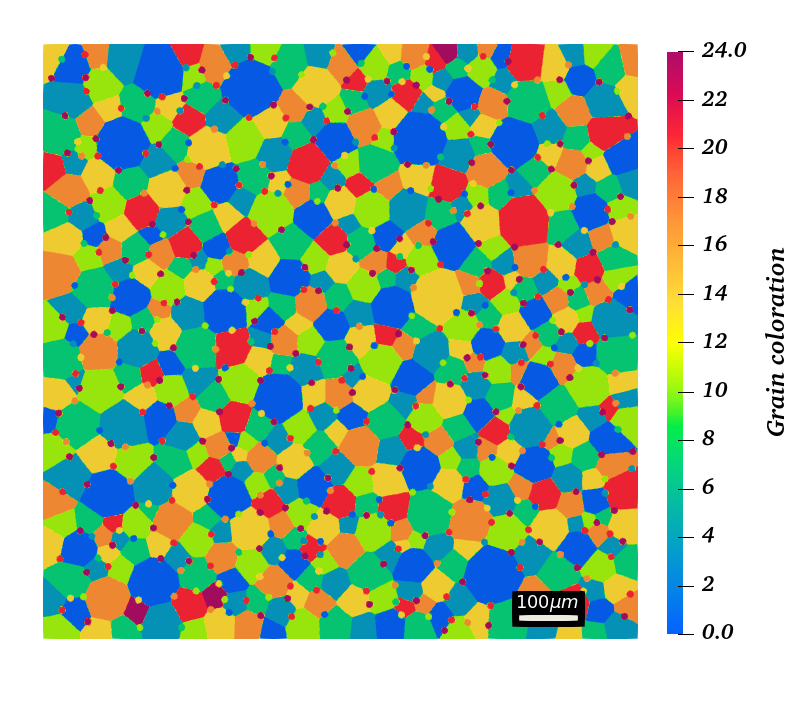}
	\captionsetup{justification=centering}
	\caption{Initial 2D two-phase polycrystal represented using the grain coloration algorithm }
	\label{2DCaseGCrep}
\end{figure}
We now consider a 2D polycrystal with a total of $492$ parent austenite grains in a square domain of side \SI{1}{\milli\meter}. The ferrite nuclei are imposed randomly along the austenite grain boundaries with an initial radius of \SI{6}{\micro\meter}. All nuclei have been imposed at the initial state in the sense of a site-saturated nucleation configuration. It should be remarked that no specific nucleation criteria have been considered for this case and that the values chosen are only illustrative.  Fig. \ref{2DCaseGCrep} describes the initial morphology of the considered two-phase polycrystal represented using the grain coloration algorithm \cite{Scholtes2015}. The color map indicates the LS-Id which means that all the grains with the same color belong initially to the same LS function. The initial conditions are considered at $T^i=$ \SI{1175}{\kelvin} with $C_\alpha^i=$ \SI{0.001167}{\wtpercent}, and $C_\gamma^i=$ \SI{0.020256}{\wtpercent}. A total of $450$ ferrite nuclei are considered which corresponds to an initial ferrite fraction of $f_\alpha^i =0.05089$. A non-isothermal phase transformation with a cooling rate of \SI{-1}{\kelvin\per\second} till the final temperature of $T^f=$ \SI{1075}{\kelvin} is imposed. In this case, since the thermal path corresponds to a global cooling of \SI{100}{\kelvin}, to reduce the potential error in the steady state predictions due to the linearization of the phase diagram, the reference states are considered in two folds:
\begin{equation}
    T^R=
    \begin{cases}
    \SI{1160}{\kelvin} & if \  \SI{1125}{\kelvin} \leq T\leq \SI{1175}{\kelvin}  \\
     \SI{1090}{\kelvin} & if \  \SI{1075}{\kelvin}\leq T< \SI{1125}{\kelvin}
    \end{cases}
    \label{refstate2Ddef}
\end{equation}
The data extracted at $T^R=$  \SI{1160}{\kelvin} are the same as the previous case (table \ref{1DTrdata}),  while at $T^R=$ \SI{1090}{\kelvin}, the ThermoCalc data are summarized in table \ref{2DTrdata2}. The expected final state has been summarized in table \ref{2DTfdata}. Mobility and interface energy are considered to be homogeneous in phase interfaces as well as the grain interfaces of both the phases ($\mu_{\alpha\gamma}=\mu_{\alpha\alpha}=\mu_{\gamma\gamma}=$ \SI{6e17}{\micro\meter\tothe{4}\per\joule\per\second}, and $\sigma_{\alpha\gamma}=\sigma_{\alpha\alpha}=\sigma_{\gamma\gamma}=$ \SI{1.0e-12}{\joule\per\micro\meter\squared}). The value for the interface energy is taken following \cite{loginova2003phase, Huang2006}. The diffuse phase interface thickness for this case is taken as $\eta=$ \SI{8}{\micro\meter} and the time step is fixed to \SI{0.01}{\second}.
\begin{table}[!htb]
\centering
 \begin{tabular}{ c | c | c | c | c | c } 
\specialrule{.2em}{.1em}{.1em}  
 $T^R$ (\SI{}{\kelvin}) & $C_{\alpha}^R$ (\SI{}{\wtpercent})& $C_{\gamma}^R$ (\SI{}{\wtpercent}) & $\Delta S$ (\SI{}{\joule\per\kelvin\per\micro\meter\cubed})& $m_\alpha^R$ (\SI{}{\kelvin\per\wtpercent}) & $m_\gamma^R$ (\SI{}{\kelvin\per\wtpercent})\\ [0.5ex] 
 \hline
 $1090$ & $0.0103363$ & $0.271627$ & $4.5486017\times10^{-13}$ & $-10161.375$ & $-255.83304$ \\ [1ex] 
\specialrule{.2em}{.1em}{.1em}
 \end{tabular}
 \captionsetup{justification=centering}
 \caption{Additional ThermoCalc data extracted at $T^R$ for $Fe-C$ \SI{0.02}{\wtpercent} for the 2D case.}
 \label{2DTrdata2}
\end{table}

\begin{table}[!htb]
\centering
 \begin{tabular}{ c | c | c | c } 
\specialrule{.2em}{.1em}{.1em}  
$T^f$ (\SI{}{\kelvin})& $C_{\alpha}^{eq}$ (\SI{}{\wtpercent})& $C_{\gamma}^{eq}$ (\SI{}{\wtpercent})& Ferrite fraction, $f_\alpha^{eq}$ \\ [0.5ex] 
 \hline
 1075 & 0.01179 & 0.33329 & 0.97669 \\ [1ex] 
\specialrule{.2em}{.1em}{.1em}
 \end{tabular}
 \captionsetup{justification=centering}
\caption{Expected steady state at $T^f$ with the final ferrite fraction estimated for an initial ferrite fraction of $0.019717$}
 \label{2DTfdata}
\end{table}

\begin{figure}[!htb]
	\centering
	\includegraphics[width=6cm]{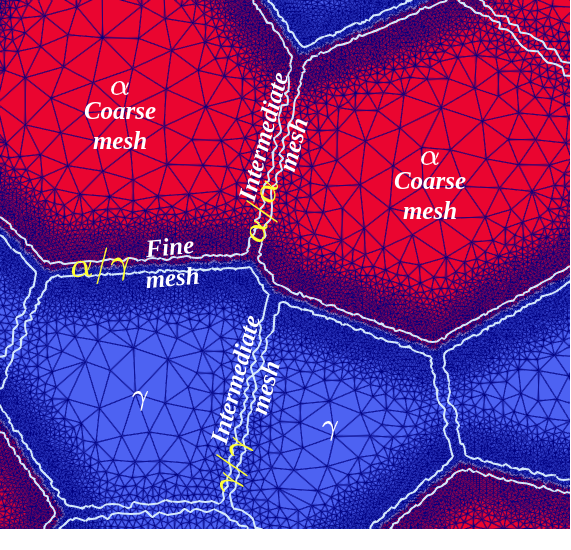}
	\captionsetup{justification=centering}
	\caption{Illustration of the adopted meshing strategy }
	\label{metintmv}
\end{figure}

A local adaptive isotropic meshing and remeshing strategy is employed \cite{Bernacki2009Rex}. A coarse mesh size is adopted in the bulk of the grains ($h_{coarse}=$ \SI{7}{\micro\meter}), whereas an intermediate mesh size ($h_{\alpha\alpha}=h_{\gamma\gamma}=$ \SI{1.2}{\micro\meter}) is used at the $\alpha/\alpha$ and $\beta/\beta$ grain interfaces and a fine mesh size ($h_{\alpha\gamma}=$ \SI{0.7}{\micro\meter}) is adopted in the phase interfaces. This strategy is illustrated in fig. \ref{metintmv} and is performed through an intersection of two different mesh metric tensors \cite{pastel-00000989} and the use of a metric-based mesher/remesher \cite{Bernacki2009Rex}. A remeshing operation is performed each forty time increments to follow the interface network migration. Such a strategy is important for saving computational time and illustrates also that the precision needed at the phase interfaces is more important than the one needed to capture only capillarity effects at the grain interfaces in each phase.

\begin{figure}[!htb]
	\centering
	\includegraphics[width=15cm]{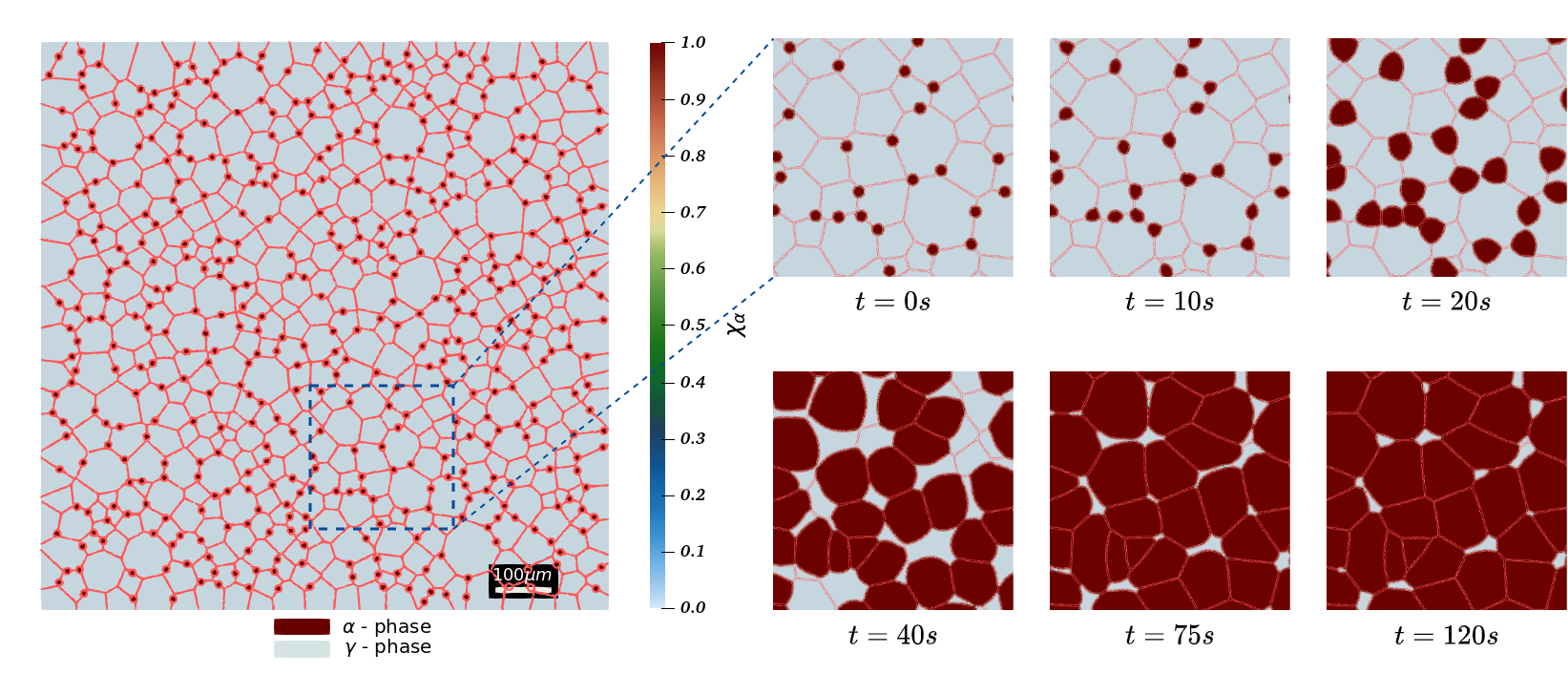}
	\captionsetup{justification=centering}
	\caption{Snapshots of phase evolution in the considered 2D two-phase polycrystal}
	\label{2DCasePhEvol}
\end{figure}

\begin{figure}[!htb]
	\centering
	\includegraphics[width=15cm]{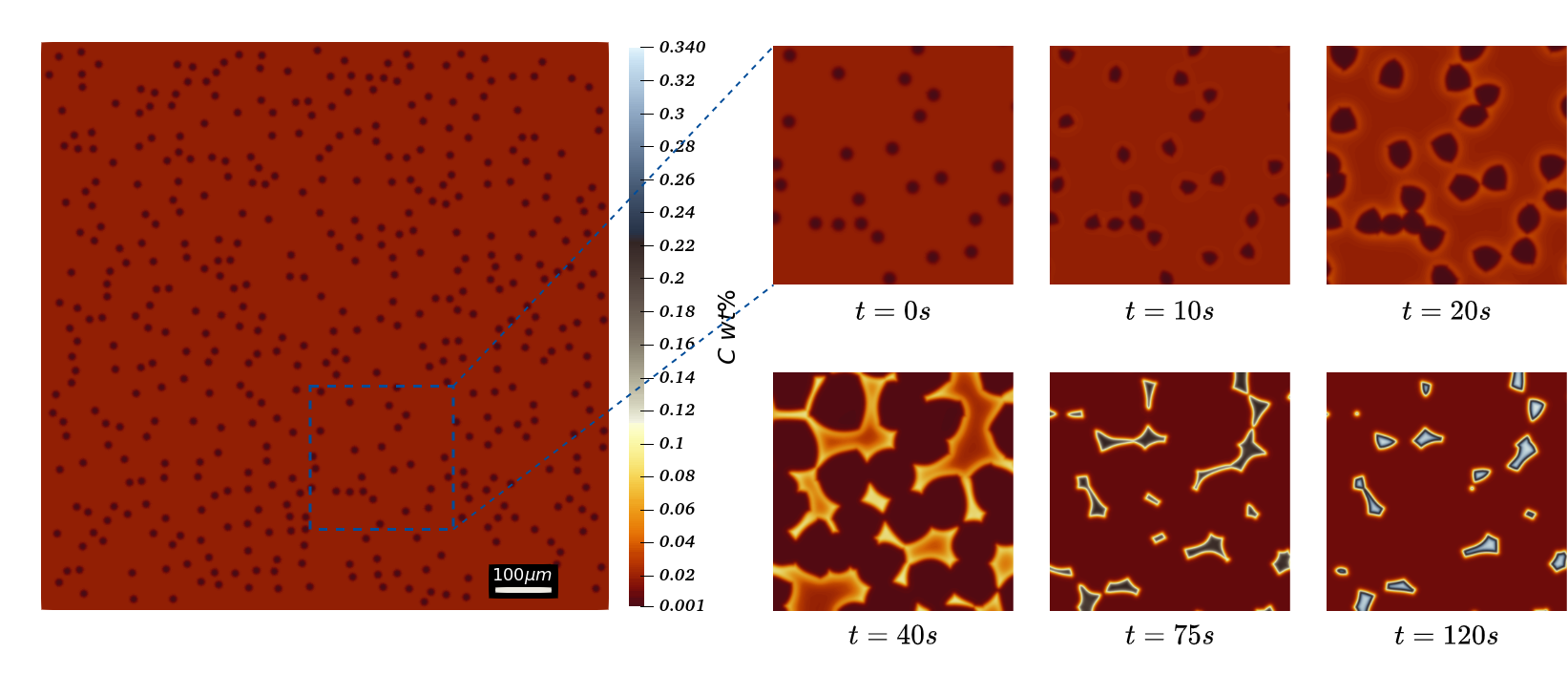}
	\captionsetup{justification=centering}
	\caption{Snapshots of solute diffusion in the considered 2D two-phase polycrystal}
	\label{2DCaseCEvol}
\end{figure}

\begin{figure}[!htb]
	\centering
	\includegraphics[width=8cm]{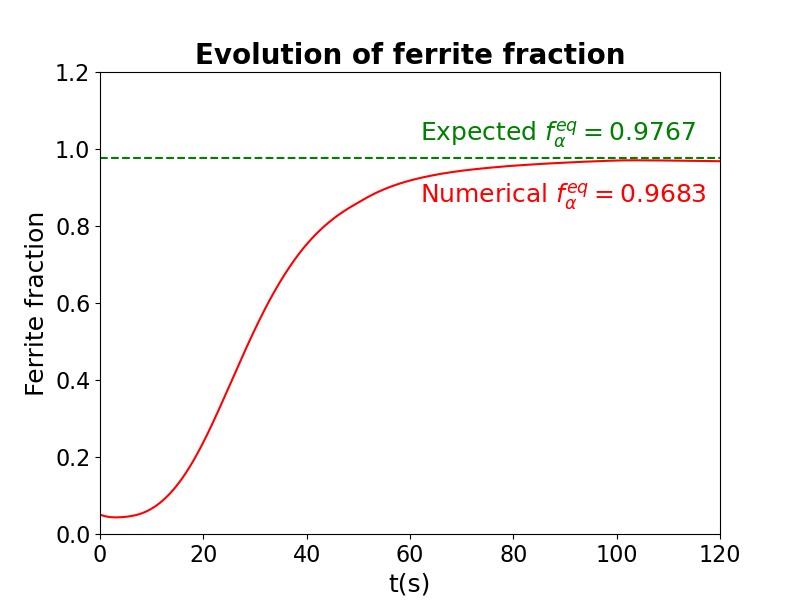}
	\captionsetup{justification=centering}
	\caption{Evolution of ferrite fraction with temperature for the 2D two-phase polycrystal case}
	\label{2DCaseFfEvol}
\end{figure}

Fig. \ref{2DCasePhEvol} illustrates the evolution of the ferrite grains at different times during the phase transformation.  Fig. \ref{2DCaseCEvol} exhibits the corresponding carbon field during the transformation. Due to the initial equilibrium, the transformation kinetics is initially slow. However, following the temperature evolution, the transformation kinetics become more important and one can observe a faster evolution of ferrite grains as indicated by the ferrite fraction curve in fig. \ref{2DCaseFfEvol}. The kinetics is then slowed down when approaching the new steady state at the final temperature. The transformation kinetics of certain clustered ferrite grains are also delayed due to the soft impingement of their diffusion fields as they continue to grow into each other. In fig. \ref{2DCaseCEvol}, one can observe carbon enrichment in the austenite grains due to the rejection of carbon from the ferrite grains. This enrichment is much more significant in those smaller austenite domains which are surrounded by several ferrite grains. Due to the high diffusivity of carbon in ferrite, the concentration distribution seems to be more homogeneous in the ferrite grains compared to that in the austenite grains.  

For the considered final state, we observe almost a complete phase transformation, with a final ferrite fraction expected to be $0.9767$ for the imposed initial grain morphology. It should be highlighted that the expected final ferrite fraction is computed analytically by applying mass conservation while not accounting for any capillarity effects. However, in the numerical simulation, capillarity effects are taken into account. So, considering this aspect and the fact that the solute mass is never perfectly conserved during the simulation, it is normal to obtain small differences between the expected and the numerically estimated final ferrite fraction. From fig. \ref{2DCaseFfEvol}, one can observe that the ferrite fraction converges towards $f_\alpha^{eq, \ num}=0.9683$ at the final state of \SI{1075}{\kelvin}. The final equilibrium concentrations are found to be $C_\alpha ^{eq, \ num}=$ \SI{0.01061}{\wtpercent}, and $C_\gamma ^{eq, \ num}=$ \SI{0.327}{\wtpercent}, which are in agreement with the ThermoCalc estimations summarized in table \ref{2DTfdata}. When performed only with one reference state at $T=$ \SI{1160}{\kelvin} (see tab. \ref{1DTrdata}), the same simulation yields a final equilibrium concentration of \SI{0.266}{\wtpercent} and \SI{0.0122}{\wtpercent} for $\gamma$ and $\alpha$ phase, respectively. These estimations are much farther from the expected values, thus vindicating the choice of using multiple reference states when the thermal paths are longer. 

\begin{figure}[!htb]
	\centering
	\captionsetup{justification=centering}
	\subfloat[Initial distribution]{\includegraphics[width=7.5cm]{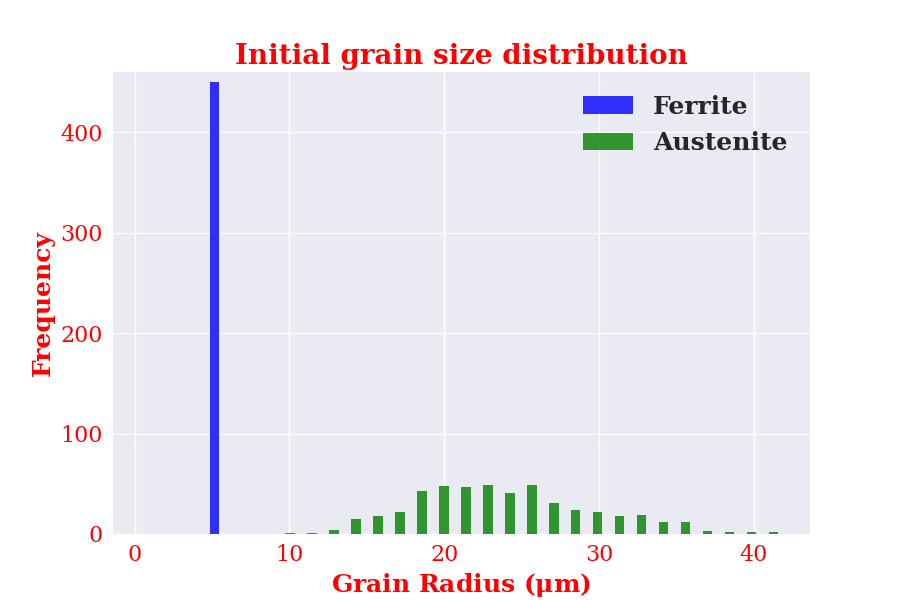}\label{2DGSDI}}
	\subfloat[Final distribution]{\includegraphics[width=7.5cm]{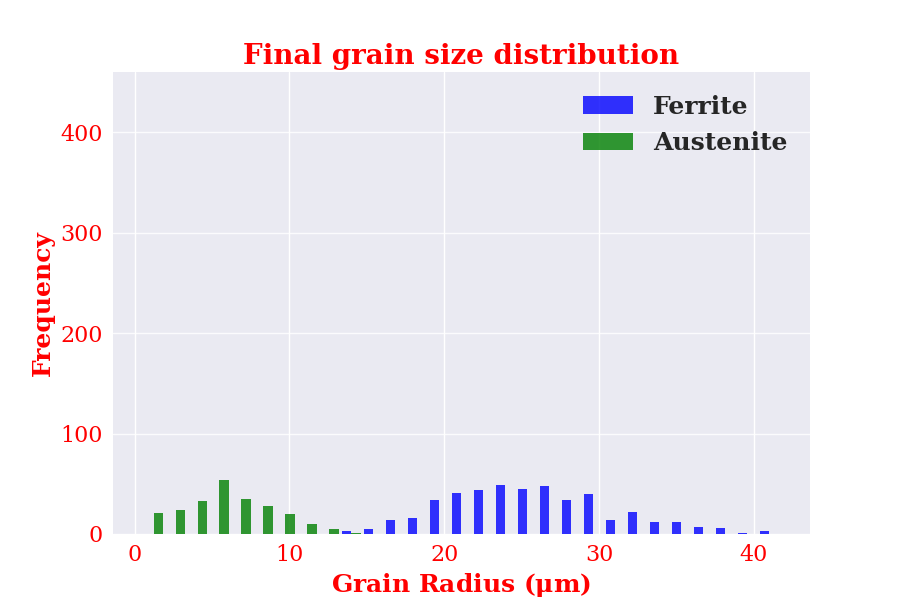}\label{2DGSDF}}
	\caption{Grain size distributions at the initial and the final state for the two phases}
	\label{2DGSD}
\end{figure}

\begin{figure}[!htb]
	\centering
	\includegraphics[width=8cm]{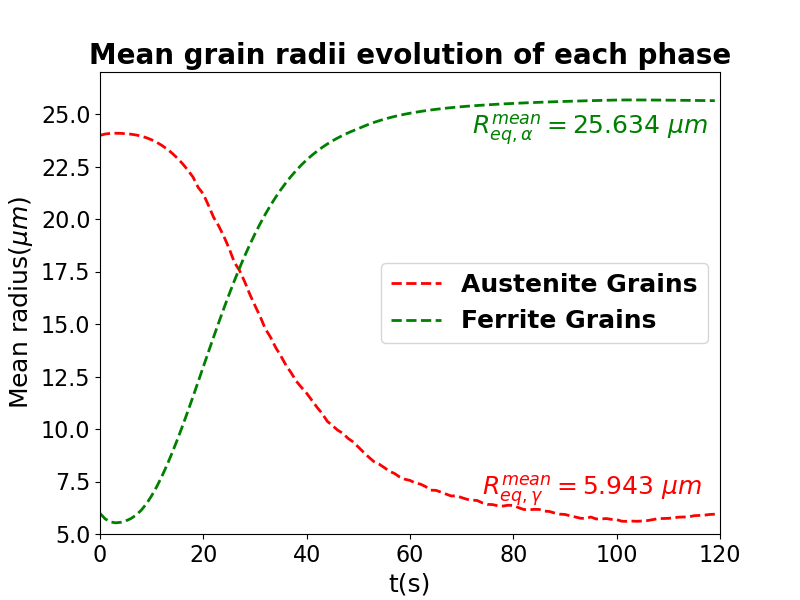}
	\captionsetup{justification=centering}
n	\caption{Mean grain radii evolution for the parent and the product phase}
	\label{2DRgevol}
\end{figure}

Figs. \ref{2DGSDI} and \ref{2DGSDF} illustrate the initial and the final grain size distribution respectively. For a given grain, its size is here defined as the radius of an equivalent circle with the same area. Fig. \ref{2DRgevol} shows the evolution of the arithmetic mean grain size for the two phases. At the initial state, one can see that all the ferrite nuclei have been imposed with the same radius of \SI{6}{\micro\meter}, while the austenite grains are normally distributed with an initial arithmetic mean grain size of around \SI{24.75}{\micro\meter}. At the final state, one can observe a bimodal normal distribution with large evolution for the two phases. The austenite grain distribution shifts to the left while the ferrite distribution shifts to the right as the austenite phase decomposes at the expense of the product ferrite phase.  From fig. \ref{2DRgevol}, one can observe that the ferrite grains converge to an arithmetic mean grain size of \SI{25.634}{\micro\meter}, while austenite grains converge to an arithmetic mean grain size of \SI{5.943}{\micro\meter}.

\pagebreak
\section{Conclusions and perspectives}
\label{concl}
A level-set (LS) based global numerical framework in a finite element context has been presented to simulate microstructural evolution in metallic two-phase polycrystalline materials. This framework has been principally illustrated in the context of diffusive solid-solid phase transformation (DSSPT). However, it has been shown that the kinetic framework presented has the potential to seamlessly take into account contributions from the stored energy due to plastic deformation as well as the grain growth effects of both the parent and the product phase. A pseudo-1D case was considered to simulate DSSPT in the context of austenite decomposition in steels. The obtained steady state characteristics were in good agreement with the ThermoCalc estimations and the obtained kinetics were in good agreement with a proposed 1D semi-analytical sharp interface model based on an extension from an existing approach. The potential of the proposed LS formulation to simulate DSSPT in a polycrystal context was illustrated through a representative 2D two-phase polycrystal case. To optimize the computational time, a specific adaptive meshing/ remeshing strategy has been employed such that the local mesh refinement is finer across the phase interfaces but relatively coarser across the grain interfaces of similar phases. For a continuous cooling phase transformation, if the thermal path is longer, it has been shown that the error in the linearization of the phase diagram for deriving the driving pressure could be minimized by considering multiple reference states. 
Since in this work, no nucleation criterion has been implemented yet, one of the perspectives is to be able to implement different criteria as a function of the cooling rate. Experimental validation of this new numerical framework in the context of austenite decomposition in steels is also planned. The current framework could also be adapted to simulate DSSPT in a multi-component system by taking into account the effects of substitutional elements on the transformation kinetics. It would also be interesting to enrich the mobility and the grain boundary energy description to be more physical concerning the anisotropy of these parameters. With minor modifications, the current framework could also be potentially adapted to simulate other diffusive solid-state phenomena such as Ostwald ripening mechanism. It would also be of interest to consider configurations with a residual stored energy field to be able to simulate phase transformation with recrystallization and grain growth. Some of these prospects will be addressed in a forthcoming publication.  

\section*{Acknowledgements}
The authors thank ArcelorMittal, Aperam, Ascometal, Aubert \& Duval, CEA, Constellium, Framatome, and Safran companies and the ANR for their financial support
through the DIGIMU consortium and RealIMotion ANR industrial Chair (Grant No. ANR-22-CHIN-0003).

\bibliographystyle{ieeetr}

\small{
	\bibliography{biblio}
}

\appendix
\newpage%
\appendixpage
\addappheadtotoc
\renewcommand{\thesection}{\Alph{section}}
\section{Mixed-mode semi-analytical 1D phase transformation model}\label{semianalytic}
\begin{figure}[!htb]
	\centering
	\captionsetup{justification=centering}
	\subfloat[Initial concentration profile at $t=$ \SI{0}{\second}]{\includegraphics[width=10cm]{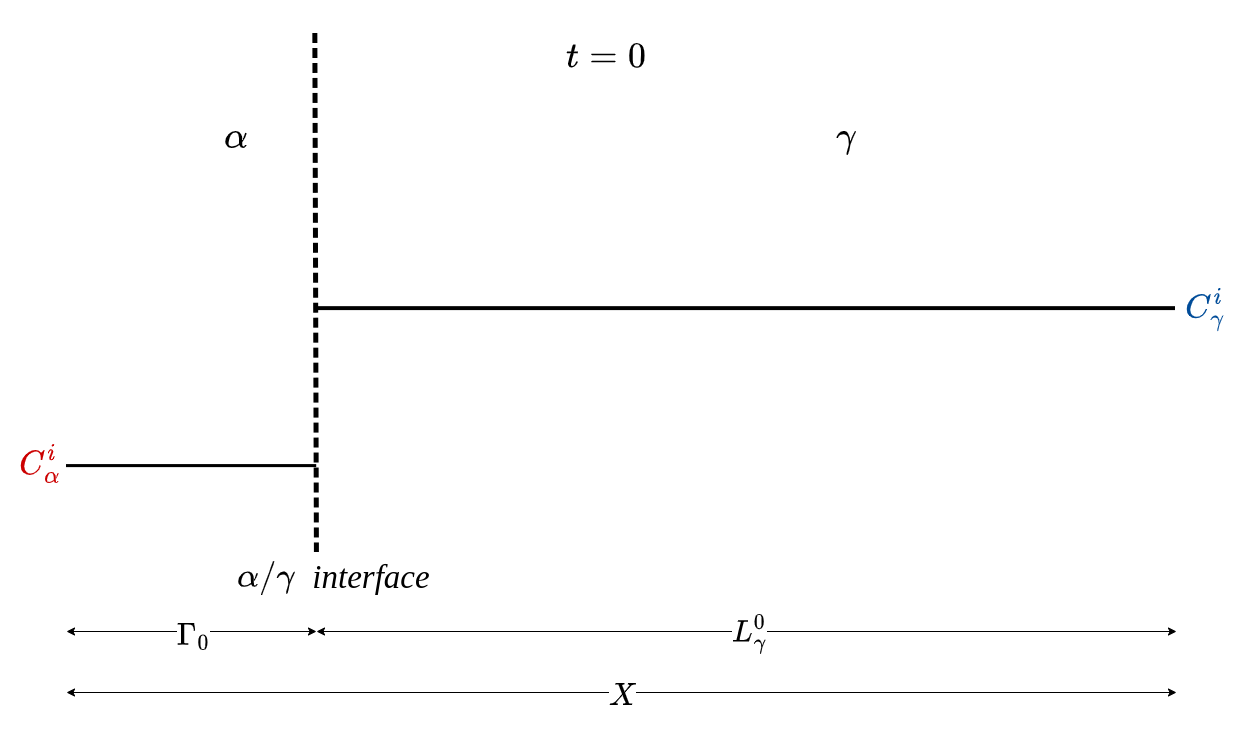}\label{SACptini}}\\
	\subfloat[Expected concentration profile at $t>$ \SI{0}{\second}]{\includegraphics[width=10cm]{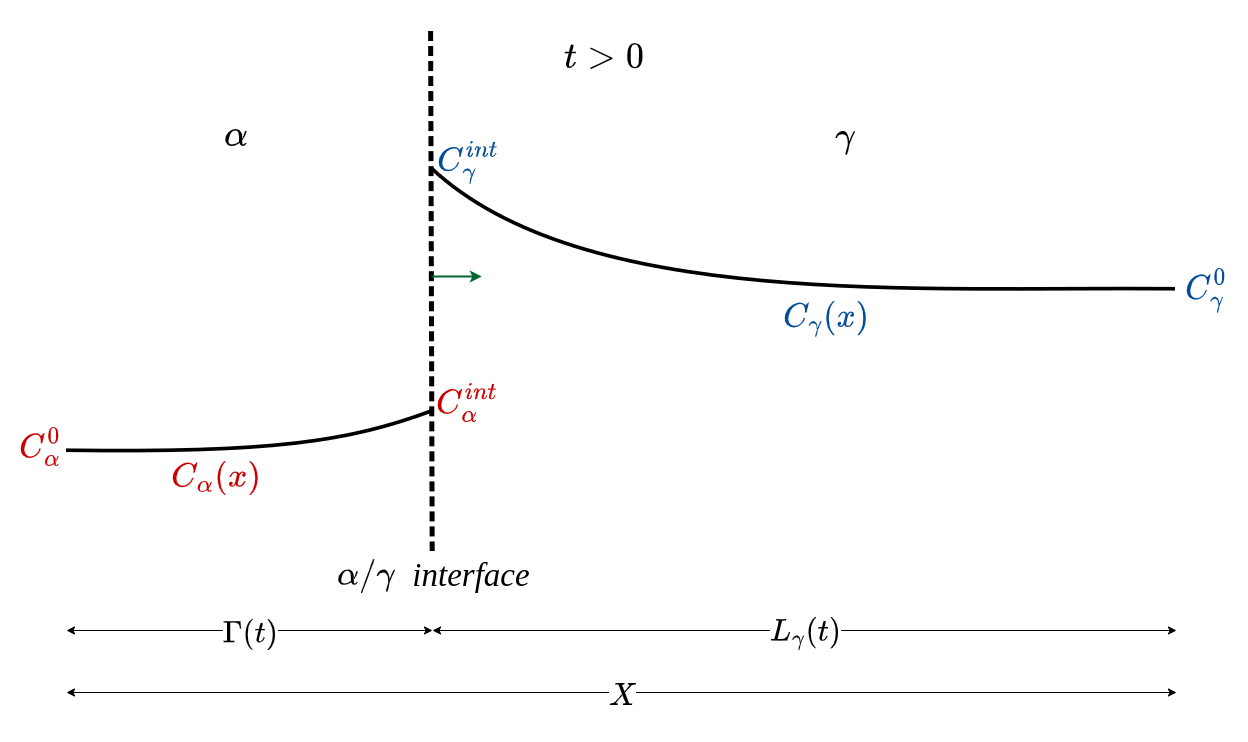}\label{SACptg0}}
	\caption{Illustration of concentration profiles expected at different instants}
	\label{SACprofdemo}
\end{figure}
As proposed in the works of Chen et al. \cite{chen2011modeling}, the idea is to consider the solute concentration profile in $\gamma$ phase ($C_\gamma(x)$) at time $t$ as a quadratic function of position ($x$):
\begin{equation}
    C_\gamma(x)=A_1 + A_2(x-\Gamma)+A_3(x-\Gamma)^2 \qquad \forall \left\{x : \Gamma(t)< x\leq X\right \} ,
    \label{CgSAprof}
\end{equation} where $A_1, A_2 \ and \ A_3$ are pre-factors, $\Gamma(t)$ is the phase interface position at time t. Chen et al. assumed diffusion to be instantaneous in the $\alpha$ phase, thus immediately
attaining the equilibrium concentration, $C_\alpha^{eq}$. This assumption is however not mandatory. Instead, one can assume that the solute concentration profile in the $\alpha$ phase follows a similar quadratic function of $x$:
\begin{equation}
    C_\alpha(x)=B_1 + B_2(x-\Gamma)+B_3(x-\Gamma)^2 \qquad \forall \left\{x : 0\leq x<\Gamma(t)\right \}.
    \label{CaSAprof}
\end{equation}
Fig. \ref{SACprofdemo} illustrates an example of concentration profiles expected in a domain of length $X$. The length of austenite side at any instant is given by $L_\gamma(t)$, such that $X = \Gamma(t) + L_\gamma(t)$. $C_\alpha^{int}$ and $C_\gamma^{int}$ are the concentrations at the sharp interface of the $\alpha$ and $\gamma$ phases respectively. Likewise, $C_\alpha^{0}$ and $C_\gamma^{0}$ are the far field concentrations in the corresponding phases. 

The pre-factors of eqs.\eqref{CaSAprof} and \eqref{CgSAprof} can be determined by applying the following boundary conditions:
\begin{equation}
    \begin{gathered}
    \begin{cases}
    C_\alpha(x=\Gamma^-) = C_\alpha^{int} & \qquad C_\gamma(x=\Gamma^+) = C_\gamma^{int} \\
    C_\alpha(x=0) = C_\alpha^{0} & \qquad  C_\gamma(x=X) = C_\gamma^{0} \\
    \left. \frac{\partial C_\alpha}{\partial x} \right|_{x=0}= 0 & \qquad  \left. \frac{\partial C_\gamma}{\partial x} \right|_{x=X}= 0
    \end{cases} \qquad \forall t>0.
    \end{gathered}
    \label{SABCs}
\end{equation}
The concentration profiles are then found to be:
\begin{equation}
    \begin{gathered}
    \begin{cases}
    C_\alpha(x)=C_\alpha^0 + \left(C_\alpha^{int} - C_\alpha^0\right)\left(\frac{x}{\Gamma}\right)^2, & \forall \left\{x : 0\leq x<\Gamma(t)\right \}\\
    C_\gamma(x)=C_\gamma^0 + \left(C_\gamma^{int} - C_\gamma^0\right)\left(1-\frac{\left(x-\Gamma\right)}{L_\gamma}\right)^2, & \forall \left\{x : \Gamma(t)< x\leq X\right \}
     \end{cases}
    \end{gathered}  \qquad \forall t>0.
    \label{SAprofeqsfinal}
\end{equation}
$L_\gamma$ serves as the width of the concentration profile ($C_\gamma$) on the austenite side. Likewise, the width of the profile ($C_\alpha$) on the ferrite side is controlled by $\Gamma$. 

Considering the boundary conditions, the solute mass needs to be conserved at any time. So, applying macroscopic solute mass balance at the time, $t>0$, we obtain:
\begin{equation}
    \int_0^{\Gamma(t)}C_\alpha(x)dx+\int_{\Gamma(t)}^{X}C_\gamma(x)dx=\int_0^{\Gamma^0}C_\alpha^idx+\int_{\Gamma^0}^{X}C_\gamma^idx.
    \label{macromassbal}
\end{equation}
By imposing that the solute concentrations at the interface redistribute at a constant ratio equal to the partitioning ratio at equilibrium (eq.\eqref{keq}) and that the far-field concentrations also respect this ratio at any instant, we obtain:
\begin{equation}
C_\alpha^{int}=kC_\gamma^{int} \quad and \quad C_\alpha^{0}=kC_\gamma^{0}, 
\label{SAhypo} 
\end{equation} with $k$ computed using eq.\eqref{keqfinal}. Expanding eq.\eqref{macromassbal} with the above hypotheses, we obtain:
\begin{equation}
C_\gamma^{0}=\frac{C_\gamma^{int}\left[\left(\Gamma^3-X^3\right)k-L_\gamma\Gamma^2\right]+3\Gamma^2\Gamma_0\left[C_\alpha^i-C_\gamma^i\right]+3C_\gamma^{i}X\Gamma^2}{2L_\gamma\Gamma^2+k\left[4\Gamma^3-X^3\right]}. 
\label{SAffCg0} 
\end{equation}

The kinetic equation for interface migration is given by:
\begin{equation}
\frac{\partial \Gamma}{\partial t} = \bm{v}\cdot\bm{n} = v_n, 
\label{SAkineqn}
\end{equation}with $v_n=\mu \Delta G$ without capillarity effects. The driving pressure, $\Delta G$ is given by the linearization of the phase diagram as already detailed in section (\ref{DGdescsection}):
\begin{equation}
\Delta G = \Delta S\left[(T^R - T) + 0.5m^R_\alpha\left(C_\alpha^{int} - C_\alpha^{R}\right) + 0.5m^R_\gamma\left(C_\gamma^{int} - C_\gamma^{R}\right) \right].
\label{DGSA}
\end{equation}
Considering no accumulation of solutes at the interface, the inward and outward solute fluxes at the interface must respect the following balance equation:
\begin{equation}
v_n\left[C_\gamma^{int}-C_\alpha^{int}\right] = \left.\llbracket \bm{J}\rrbracket\right|_\Gamma \cdot\bm{n} = D_\alpha^C\left.\frac{\partial C_\alpha}{\partial x}\right|_\Gamma-D_\gamma^C\left.\frac{\partial C_\gamma}{\partial x}\right|_\Gamma. 
\label{intfluxbal}
\end{equation}
Further expanding and making necessary substitutions, one can rewrite this equation as:
\begin{equation}
f(C_\gamma^{int})=\mu\Delta GC_\gamma^{int}\left(1-k\right) -  \frac{2D_\alpha^Ck}{\Gamma}\left(C_\gamma^{int}-C_\gamma^{0}\right)-\frac{2D_\gamma^C}{L_\gamma}\left(C_\gamma^{int}-C_\gamma^{0}\right)=0.
\label{cgintnonlineq}
\end{equation}
Since $\Delta G$, and $C_\gamma^0$ are both functions of $C_\gamma^{int}$, the above eq.\eqref{cgintnonlineq} is a non-linear equation.
\subsection*{Resolution procedure:}
\begin{itemize}
    \item Eq.\eqref{cgintnonlineq} is resolved iteratively for $f(C_\gamma^{int})=0$ to compute $C_\gamma^{int}$.
    \item From $C_\gamma^{int}$; $C_\alpha^{int}$, and $v_n$ are computed.
    \item The interface is then migrated using eq.\eqref{SAkineqn} with an explicit Euler scheme: \begin{equation*}
        \Gamma^{n+1}=\Gamma^{n}+v_n\Delta t,
    \end{equation*}where $n$ is the index for time stepping and $\Delta t$ is the chosen time step. $L_\gamma^{n+1}=X-\Gamma^{n+1}$ can then be computed.
    \item $C_\gamma^0$, $C_\alpha^0$, and the concentration profiles $C_\alpha (x)$ and $C_\gamma (x)$ at time $t^{n+1}=(n+1)\Delta t$ are then computed
\end{itemize}



\end{document}